\documentclass[aps,prfluids,floatfix,superscriptaddress,longbibliography]{revtex4-2}

\usepackage[british,UKenglish]{babel}
\usepackage{csquotes}
\usepackage{nicefrac}
\usepackage{xspace}
\usepackage{bm}
\usepackage{amsmath}
\usepackage{mathtools}
\usepackage{amssymb,amsmath,amsfonts,mathrsfs,bbm}
\usepackage[bbgreekl]{mathbbol}
\usepackage{amsthm}
\usepackage{braket}
\usepackage{cases}
\usepackage{empheq}
\usepackage{upgreek}
\usepackage{tabularx}
\usepackage{booktabs}
\usepackage{array}
\usepackage{enumerate}
\usepackage{enumitem}
\usepackage{color}
\usepackage[dvipsnames]{xcolor}
\usepackage{pgfplots}
\usepackage{tikz}
\usepackage{multirow}
\usetikzlibrary{plotmarks}
\usepackage[colorlinks=true]{hyperref}
\hypersetup{
	linkcolor = {blue},
	citecolor = {blue}
}

\usepackage{graphicx}
\usepackage{subfig}
\usepackage[justification=raggedright, singlelinecheck=false]{caption}

\usepackage[toc,page]{appendix} 

\newtheoremstyle{break}
{\topsep}{\topsep}%
{\itshape}{}%
{\bfseries}{}%
{\newline}{}%
\theoremstyle{break}

\newcolumntype{P}[1]{>{\centering\arraybackslash}p{#1}}

\newcommand{\ie}{\textit{i.e.}\ }

\newcommand{\eg}{\textit{e.g.}\ }

\definecolor{ballblue}{rgb}{0.13, 0.67, 0.8}
\definecolor{azure}{rgb}{0.0, 0.5, 1.0}
\newcommand{\Rep}[1]{\textcolor{black}{#1}}

\newcommand{\dd}{\mathrm{d}}

\newcommand{\Pdf}{\mathcal{P}}

\newcommand{\tauO}{\tau_{\omega}}
\newcommand{\tauI}{\tau_{\mathrm{I}}}
\newcommand{\s}{\sigma}

\newcommand{\Ku}{\mathrm{Ku}}
\newcommand{\nicepi}{\nicefrac{\pi}{2}}

\newcommand{\Ito}{It\^o\xspace}
\newcommand{\Stra}{Stratonovich\xspace}
\newcommand{\OU}{Ornstein--Uhlenbeck\xspace}

\newcommand{\FP}{Fokker--Planck\xspace}

\newcommand{\C}{\mathcal{C}}

\newcommand{\D}{\mathcal{D}}

\newcommand{\ha}{\widehat{a}}

\newcommand{\Diso}{\mathcal{D}^{\textrm{iso}}}
\newcommand{\Daniso}{\mathcal{D}^{\textrm{aniso}}}
\newcommand{\Dint}{\mathcal{D}^{\textrm{int}}}
\newcommand{\DD}{{\mathscr{D}_{\sigma^*}}}
\newcommand{\aliso}{\alpha_{\textrm{iso}}}
\newcommand{\alaniso}{\alpha_{\textrm{aniso}}}
\newcommand{\alint}{\alpha_{\textrm{int}}}

\newcommand{\brmX}{\bm{\mathrm{X}}}
\newcommand{\w}{\textrm{w}}

\newcommand{\pst}{ \mathcal{P}_{\rm st} }

\newcommand{\btheta}{{\bar{\theta}}}

\newcommand{\Pst}{\mathcal{P}_{\rm st}}

\newcommand{\dt}{\partial_t}

\newcommand{\B}{\mathsf{B}}

\newcommand{\RN}[1]{%
	\textup{\uppercase\expandafter{\romannumeral#1}}%
}

\newcommand\smallO{
	\mathchoice
	{{\scriptstyle\mathcal{O}}}
	{{\scriptstyle\mathcal{O}}}
	{{\scriptscriptstyle\mathcal{O}}}
	{\scalebox{.7}{$\scriptscriptstyle\mathcal{O}$}}
}

\newcommand{\cemef}{MINES Paris, PSL Research University, CNRS, Cemef, Sophia-Antipolis, France}
\newcommand{\inria}{Universit\'e C\^ote d'Azur, Inria, CNRS, Cemef, Sophia-Antipolis, France}

\begin{document}

\title{Stochastic model for the alignment and tumbling of rigid fibres \\
		in two-dimensional turbulent shear flow}

\author{Lorenzo Campana} \affiliation{\inria}
\author{Mireille Bossy}  \affiliation{\inria}
\author{J\'er\'emie Bec} \affiliation{\inria}\affiliation{\cemef} 

\begin{abstract}
Non-spherical particles transported by an anisotropic turbulent flow preferentially align with the 
mean shear and intermittently tumble when the local strain fluctuates. Such an intricate behaviour is 
here studied for small, inertialess, rod-shaped particles embedded in a two-dimensional turbulent 
flow with homogeneous shear. A Lagrangian stochastic model for the rods angular dynamics is 
introduced and compared to the results of direct numerical simulations.  The model consists in 
superposing a short-correlated random component to the steady large-scale mean shear, and can 
thereby be integrated analytically.  
Reproducing the single-time orientation statistics obtained numerically requires to take account of 
the mean shear, of anisotropic velocity gradient fluctuations, and of the presence of persistent 
rotating structures that combine together to bias cumulative Lagrangian statistics.
The model is then used to address two-time statistics. The notion of 
tumbling rate is extended to diffusive dynamics by introducing the stationary probability flux of the 
rods unfolded angle, which provides information on the overall, cumulated rotation of the particle.  
The model is found to reproduce the long-term effects of an average shear on 
the mean and the variance of the fibres angular increment.
Still, for intermediate times, the model fails catching violent fluctuations of the rods rotation that are 
due to trapping events in coherent, long-living eddies.
\end{abstract}

\maketitle

\section{Introduction}\label{chap:intro}
The control and prediction of flows seeded with non-spherical particles (fibres, discs, or inclusions 
with more general shapes) are important in many industrial and natural processes. In papermaking, 
mechanical properties are regulated by the alignment of cellulose fibres in the 
pulp~\cite{lundell2011fluid}. The shape and orientation of fractal soot emitted by combustion 
engines determines their radiative properties as aerosols~\cite{moffet2009situ}. Particles 
non-spherical shape is key to study the dispersion of pollen and  
seeds~\cite{sabban2017temporally}, the lifecycle of diatom plankton~\cite{musielak2009nutrient} 
and sediment transport in rivers~\cite{vercruysse2017suspended}. Besides, the addition of fibres in 
a fluid flow can significantly alter the suspension 
rheology~\cite{butler2018microstructural,daghooghi2015influence}.

In most of these applications, the flow is highly turbulent and the particles rotational dynamics, their 
alignment and correlations with the flow become of considerable interest. When their size falls in 
the active turbulent scales, the particles filter the fluid in a complex manner and display a very 
intricate dynamics. Particles with sizes in the dissipative range have a more tractable behaviour. 
When their slip velocity is small enough, the local flow has a weak inertia and Stokes solutions can 
be used to relate the particles translational and rotational dynamics to the local velocity field and its 
gradient tensor. This approach was introduced by~\citet{jeffery1922motion} and used to investigate various interesting phenomena, such as the tumbling of rods placed in a constant shear flow and subject to thermal fluctuations, and their resulting  periodic motions on closed (Jeffery's) orbits. 

In turbulence, fluctuations originate from the small-scale motions of the flow and this affects the 
angular dynamics of anisotropic particles. Such questions have recently been the subject of a 
renewed interest. At an experimental level, particle tracking techniques allowed to reconstruct 
particle orientational dynamics in several turbulent water flows~\cite{parsa2011rotation, 
bellani2012shape, parsa2014inertial, marcus2014measurements, bounoua2018tumbling}. 
As to numerical studies, they consist in simulating, in addition to the fluid flow, the orientation of 
particles by integrating Jeffery's equation along Lagrangian trajectories. Simulations have been 
carried out in homogeneous isotropic turbulence~\cite{pumir2011orientation,parsa2012rotation}, as 
well as in turbulent channel flows~\cite{mortensen2008dynamics}, in two-dimensional 
convection~\cite{calzavarini2020anisotropic}  and in chaotic velocity 
fields~\cite{wilkinson2009fingerprints}. From a theoretical perspective, most studies 
consisted in deriving model equations for the probability distribution of orientations, in which 
turbulent fluctuations are approximated by an effective isotropic diffusion 
term~\cite{shin2005rotational}.  We refer the reader to~\cite{voth2017anisotropic} for a recent 
review of these results. Despite several studies in turbulent channel flows, much needs to be 
understood in the presence of flow anisotropies. 

We here focus on the statistics of alignment and tumbling in the presence of a mean shear, for which 
we expect spheres and rods to rotate in qualitatively different fashions. When the particle's Reynolds 
number is small, the local flow is well approximated by a Stokes flow. If in 
addition their inertia can be neglected, spheres will rotate with an angular velocity given by half the 
flow vorticity. Studying the dynamics of ellipsoidal particles in turbulence is a challenging problem 
because it requires to understand how the Lagrangian statistics of velocity gradients influence the 
orientation dynamics of particles. A number of recent numerical and theoretical works studied the 
orientation statistics of non-spherical particles (rods, ellipsoids, disks) by assuming that the velocity 
field is isotropic and Gaussian, \eg results from the superposition of linear \OU 
processes~\cite{brunk1998turbulent, pumir2011orientation, vincenzi2013orientation}. This 
assumption, albeit restrictive, allows for a fully analytical expression of the probability density 
function of the particles orientation. More refined models for Lagrangian velocity gradients have 
been proposed~\cite{girimaji1990diffusion, chertkov1999lagrangian, chevillard2006lagrangian, 
biferale2007multiscale}, with the aim to represent specific features encoded in the tensor, such as 
the alignment of vorticity with the strain-rate eigenvectors, the rate of deformation and shape of 
fluid material volumes, non-Gaussian statistics or intermittency. \citet{chevillard2013orientation} 
have studied the orientation of tri-axial ellipsoids in direct numerical simulations of homogeneous 
isotropic turbulence and compared their results to Lagrangian stochastic models based on the 
recent fluid deformation approximation. This approach, which includes realistic strain--vorticity 
correlations, was moreover used in~\cite{pereira2018multifractal} to develop a stochastic model that 
accounts for the fluid velocity intermittency. As stressed in~\cite{chevillard2013orientation, 
pereira2018multifractal}, coupling these models to the orientation dynamics of non-spherical 
particles turns out to be a precise and demanding way to assess their accuracy. However, in 
contrast with the homogeneous and isotropic case, only few analytical results exist on the probability 
distribution of orientations in flows with a mean shear. The orientation dynamics of rod-like polymers 
was studied numerically~\cite{celani2005polymers} and analytically~\cite{turitsyn2007polymer} 
using the superposition of a mean shear and isotropic fluctuations with short time correlations. The 
dynamics of semi-flexible objects in an extensional flow was otherwise analysed 
in~\cite{plan2016tumbling,henry2018tumbling}.

In this article, we rely on the use of direct numerical simulation (DNS) to validate the development of 
a Lagrangian stochastic model. Our model consists in approximating the flow viewed by the rods 
as the superposition of a constant shear with a random component corresponding to a chaotic, 
fluctuating velocity field. In the spirit of classical approaches~\cite{batchelor1959small, 
kraichnan1968rh}, the fluctuating part is approximated as a Gaussian white-in-time noise with 
prescribed correlations. The assumption of temporal decorrelation is adequate when the correlation 
time of the flow is short compared to the time scales of relevance for the rod evolution. Furthermore, 
such models are a great simplification of real flows and have been successfully applied for analysing 
transport by turbulent flows~\cite{falkovich2001particles}. This approach, which is apparently 
restrictive, has the great advantage to permit analytical derivations. As an instance, we obtain an 
exact expression for the probability distribution of the rod orientation angle.

The model that we introduce depends on a non-dimensional parameter, the Kubo number  
$\Ku=\tauI/\tau_{\omega}$, defined as the ratio between the integral correlation time of the 
Lagrangian velocity gradient and the timescale obtained from the inverse of its standard deviation. 
We compare three different models for the effective correlation tensor: The first assumes that the 
correlations are isotropic without any reference to $\Ku$; The second introduces anisotropies by 
directly measuring the instantaneous correlations of the velocity gradient components and requires 
prescribing the value of $\Ku$; The third is based on measurements from the DNS of 
orientation-dependent integral correlation times and provides the model with an effective 
anisotropic correlation tensor.  All three models reproduce qualitatively the preferential alignment of 
rods with the direction of shear, with an angular distribution that gets more peaked with 
increasing mean shear. Results show, however, that a good quantitative agreement with DNS is 
obtained only with the third approach, underlining the importance of accurately reproducing the 
anisotropies of fluctuations. 

Finally, we study the statistics of rod tumbling and of the particles angular velocity. This notion is 
ill-defined in stochastic models, and we therefore introduce an alternative way to characterise 
tumbling through the two-time statistics of particles orientation. It relies on estimating the time 
derivative of the average unfolded angular displacement, defined to account for the cumulative 
rotation of the particles. The analytic results obtained from our 
model are compared with DNS measurements, revealing that the model is in quantitative agreement 
for adequate values of the calibration parameter. However, we point out the limit of Gaussian models 
that are unable to properly reproduce large fluctuations of the rods angular displacement. Indeed, 
we observe in DNS that the probability distribution of the orientation increment displays intermediate 
algebraic tails that are a signature of trapping by the long-living vortical structures of the flow.

Our presentation is organised as follows. In Sec.~\ref{sec:DNS} we formulate the dynamics of rod-like particles in two-dimensional turbulent flow with homogeneous shear, describe the numerical method used for direct numerical simulation (DNS), and present results on the preferential alignment of rods. Section~\ref{sec:orientation_model} then introduces the Lagrangian stochastic model for the rods' orientation and analytical results for the stationary distribution of the orientation angle. The various choices for the effective gradient correlations are discussed and compared to the results of DNS. Section~\ref{sec:tumbling} is devoted to two-time statistics and provides a new definition of tumbling rate, which is investigated both numerically in DNS and analytically for the Lagrangian model. Conclusions and open questions are finally drawn in Sec.~\ref{sec:conclusion}.

\section{Settings and direct numerical simulations}\label{sec:DNS}

\subsection{Two-dimensional turbulence with homogeneous shear}
\label{subsec:shear2d}

The fluid velocity $\bm v$ is here assumed to solve the two-dimensional incompressible 
Navier--Stokes equations
\begin{equation}
	\begin{aligned}
		\partial_t\bm v + \bm v\cdot\nabla\bm v &=
		-\nabla p + \nu\nabla^2 \bm v
		-\alpha\,(\bm v-\sigma\,y\,\hat{\bm x})
		+\bm f, \\
		\nabla\cdot\bm v &= 0,
	\end{aligned}
	\label{eq:ns}
\end{equation}
where the fluid has a constant density equal to unity, the pressure $p$ enforces incompressibility, 
and $\nu$ is the kinematic viscosity. The flow experiences a drag with coefficient $\alpha$ to an 
underlying flow in the $x$ direction that varies linearly along the $y$ axis with shear rate $\sigma$.  
In the absence of external forcing ($\bm f = 0$), the fluid velocity relaxes to the two-dimensional 
shear flow $\bm v_\infty = \sigma\,y\,\hat{\bm x}$, which is a stable stationary solution to 
Eq.~\eqref{eq:ns}. In order to maintain a developed turbulent state, an input of kinetic energy is 
provided by the stochastic forcing $\bm f$, which is assumed Gaussian with zero mean, 
homogeneous, isotropic, white in time and concentrated over the large spatial scales. The 
incompressible turbulent fluctuations $\bm u = \bm v-\bm v_\infty$ are then homogeneous in space 
and stationary in time. The equations of motion are supplemented with appropriate periodic 
boundary conditions on a square domain of size $L^2$. They read $\bm u(x+L,y) = \bm u(x,y)$ and 
$\bm u(x,y+L) = \bm u(x-t\,\sigma\,L,y)$. 

We perform direct numerical simulations by using a pseudo-spectral solver. To construct periodic 
solutions that account for the mean flow, we  follow~\cite{rogallo1981numerical, 
pumir1996turbulence} and integrate the dynamics of vorticity fluctuation $\omega = \nabla \times 
\bm u = \partial_x u_y - \partial_y u_x$ on a distorting frame defined by $x' = x-t \, \s \,y$, $y'=y$. 
The integration domain is the two-dimensional torus $[0,2\pi]^2$ at resolution $128^2$. The 
distorted grid is regularly shifted back to the Cartesian grid at times multiple of $1/\s$. We make use 
of the vorticity formulation, together with using the Biot--Savard law to obtain the fluctuating 
velocity $\bm u$ as a function of the vorticity $\omega$. Time marching uses a second-order 
Runge--Kutta method, which is explicit for the non-linear term and implicit for the friction and 
viscous terms. Furthermore, simulations are performed with hyperviscosity and hypofriction in place 
of the viscous dissipation and linear friction terms appearing in the right-hand side of 
Eq.~\eqref{eq:ns}. The latter two terms are replaced with $(-1)^{p+1} \nu_p\nabla^{2p} \bm v$ and 
$(-1)^{q+1} \alpha_q\nabla^{-2q} \bm v$, respectively. The use of hyperviscosity ($p > 1$) and 
hypofriction ($q > 0$) is motivated by the resulting reduction of the scale range over which 
dissipative terms contribute substantially, whereby extending the inertial range for a given spatial 
resolution~\cite{lindborg2000kinetic,haugen2004inertial}. It has been observed that such a modified 
dissipation might affect velocity statistics at the transition between inertial and dissipative 
scales~\cite{frisch2008hyperviscosity}. Still, the situations and effects that 
we consider here are related to the direct cascade of enstrophy, which is only very weakly perturbed.
\begin{figure}[htp]
\captionsetup[subfigure]{position=bottom, labelfont=bf,textfont=normalfont,	singlelinecheck=false,  justification=centering}
\centering
\subfloat[\label{fig:stat_fluids_a}]{
		\includegraphics[scale=0.5]{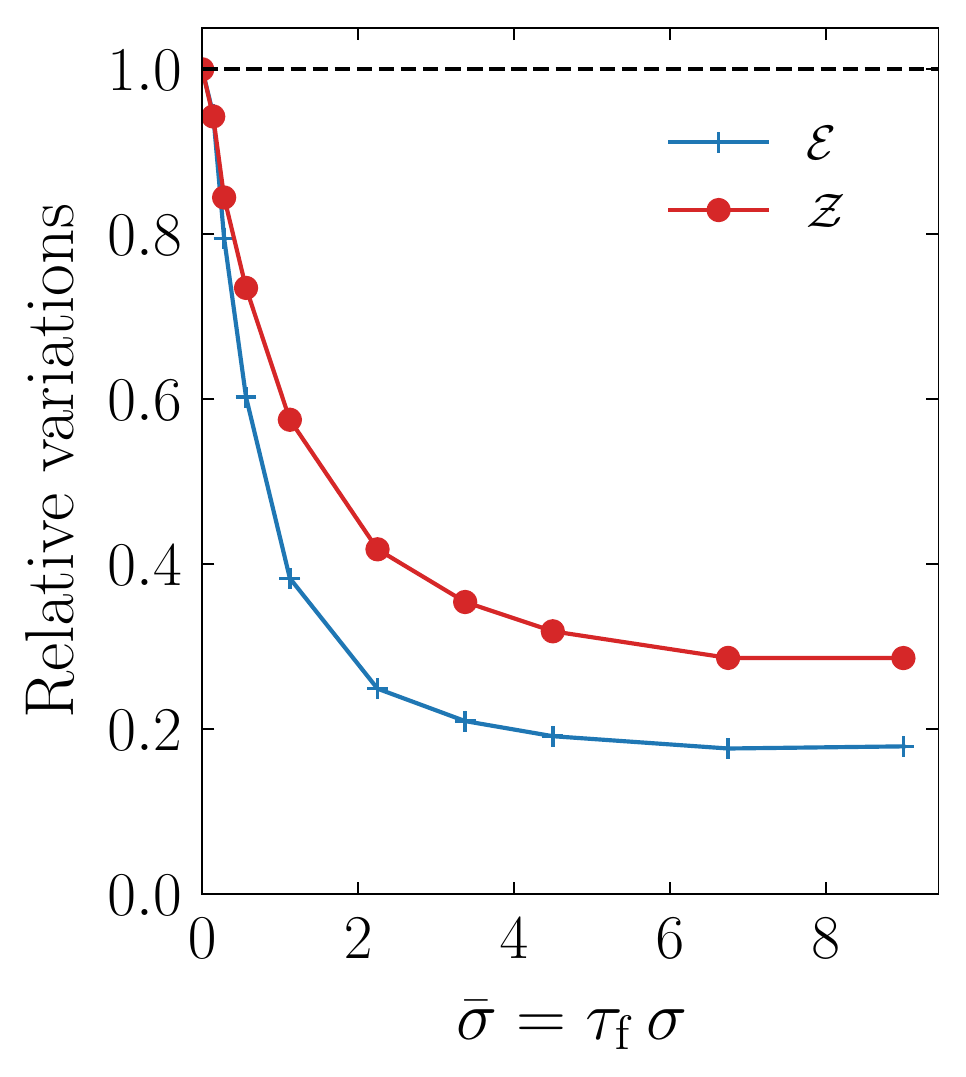}}\hfil 
\subfloat[\label{fig:stat_fluids_b}]{
		\includegraphics[scale=0.5]{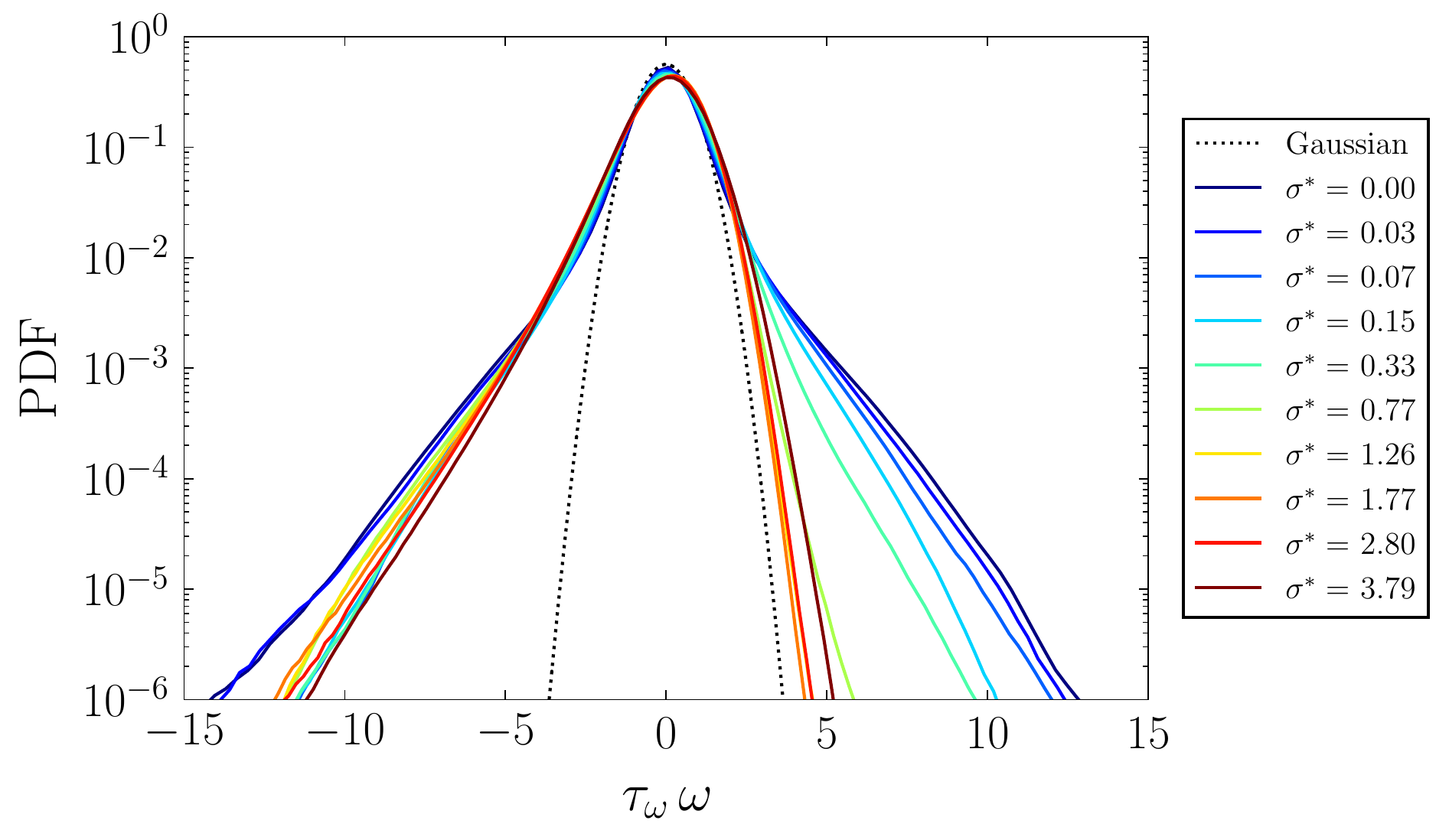}}		
\caption{(Colour online) \label{fig:stat_fluids} 
	\Rep{\protect\subref{fig:stat_fluids_a}} Relative evolution of the turbulent kinetic energy 
	$\mathcal{E} = \braket{|\bm u|^2}/2$ and enstrophy $\mathcal{Z} = \braket{\omega^2}/2$ as a 
	function of the dimensionless shear parameter $\bar{\s} = \tau_{\mathrm{f}} \, \s$. 
	\Rep{\protect\subref{fig:stat_fluids_b}} Probability density function (PDF) of the vorticity 
	$\omega$ for various values of the shear rate parameter $\s^*= \tauO \, \s$.}
\end{figure}

The power spectrum of the external force is chosen with a Gaussian shape, centred at wavenumber 
$k_{f}=4$ and with variance $\sigma_f=0.5$. The choice of an additive, white-in-time forcing 
prescribes  the rates $\varepsilon_{\rm I}$ and $\eta_{\rm I}$ at which energy and enstrophy are 
injected in the flow~\cite{boffetta2012two}. In particular, we performed several numerical  
experiments varying the shear rate $\s$ while keeping constant these two injection rates and so, the 
resulting forcing length scale $\ell_{\rm f} = (\varepsilon_{\rm I}/\eta_{\rm I})^{1/2}$ and time scale 
$\tau_{\rm f} = \eta_{\rm I}^{-1/3}$. The influence of the mean shear is then naturally weighed by the 
non-dimensional parameter $\bar{\s} = \tau_{\rm f}\,\s$. As seen in Fig.~\ref{fig:stat_fluids}\protect\subref{fig:stat_fluids_a}, 
shear has a strong effect on the energy and enstrophy of turbulent fluctuations: These two global 
quantities decrease with increasing $\bar{\s}$, evidencing a significant reduction of fluctuations.  As 
a consequence, the typical dynamical timescale of the direct cascade $\tauO = 
\braket{\omega^2}^{-1/2}$ significantly increases with shear. To account for this trend,  we 
introduce another dimensionless shear rate parameter defined as $\s^*= \tauO \, \s$.

Besides modifying global budgets, the presence of a mean shear is responsible for the development 
of anisotropies in the flow. This is evident from the one-point, one-time probability density function 
(PDF) of the vorticity $\omega$.  As can be seen in Fig.~\ref{fig:stat_fluids}\protect\subref{fig:stat_fluids_b}. Even if the 
mean value remains zero, shear strongly depletes the distribution of positive (anticyclonic) values, 
so that vorticity becomes strongly skewed toward negative (cyclonic) values when $\s^*$ increase. 
At $\s^*=0$, the distribution consists of a Gaussian core, followed by exponential tails, as predicted 
for instance in~\cite{falkovich2011vorticity}.  Shear tends to slightly shift the 
Gaussian core toward positive values. The far exponential tails persist but become skewed. While 
the negative tail is slightly depleted, the positive one is significantly affected by shear with a decay 
rate that strongly increases as a function of $\s^*$. The main effect of shear is thus to bias vortex 
filaments toward positive vorticities (Gaussian core) and to deplete anticyclonic vortical coherent 
structures (right tail).

\subsection{Rods dynamics and preferential alignment}
\label{subsec:rods_dynamics}

The particles in question are assumed to be inertialess, neutrally-buoyant, rod-shaped, and 
much smaller than the smallest active scales of the flow.  Hence their centres of mass approximately 
evolve as tracers $\dd \brmX(t)/\dd t = \bm v(\brmX(t),t)$. In addition, the particles are assumed to 
be sufficiently dilute to neglect both their interactions and their feedback onto the flow. The 
orientation of such a particle, which is specified by a unit vector $\bm p$, follows Jeffery's 
equation~\cite{jeffery1922motion} for inertialess ellipsoidal particles with an infinite aspect ratio
\begin{equation} \label{eq:jeffery_rod}
	\frac{\rm d}{\rm d t} \bm p = \mathbb{A}\, \bm p - (\bm p^{\sf T}\mathbb{A}\,\bm p)\, \bm p,
\end{equation}
where $\mathbb{A}(t)$ denotes the gradient tensor of the fluid-velocity field $\bm v$, 
evaluated at the particle's position. Its components read ${\sf A}_{ij}(t)=(\partial v_i / \partial 
x_j)(\brmX(t),t)$.  In two dimensions, Jeffery's equation~\eqref{eq:jeffery_rod} can be conveniently 
rewritten in term of  the orientation angle $\theta_t$, defined as $\bm 
p(t)=(\cos\theta_t,\sin\theta_t)$, namely
\begin{equation}
	\frac{\dd \theta_t}{\dd t} = \; \frac{\sigma}{2} \left( \cos(2\theta_t) -1 \right)+
	\frac{1}{2}\omega - \partial_x u_x\,\sin(2\theta_t)
	+\frac{1}{2}(\partial_x u_y + \partial_y u_x)\,\cos(2\theta_t),\label{eq:orientation}
\end{equation}
where the vorticity $\omega$ and $\partial_{x,y}u$ are valued at $ (\brmX(t),t)$. 
Using this representation actually provides additional information compared to a simple integration of~\eqref{eq:jeffery_rod}. It indeed allows keeping track of the full unfolded angle $\theta_t\in\mathbb{R}$  rather than limiting our analysis to its folded image $\bar{\theta}_t=\arctan(p_y/p_x)\in [-\nicepi, \nicepi]$. As we will later see, this is particularly convenient when evaluating by how many turns the orientation vector $\bm p$ has evolved over long time lags.
In our simulations, we uniformly seed the flow with tracer particles and integrate 
Eq.~\eqref{eq:orientation} along their trajectories with an initial orientation uniformly distributed over 
$[0,2\pi]$.

\begin{figure}[!h]
\captionsetup[subfigure]{position=bottom, labelfont=bf,textfont=normalfont,
	singlelinecheck=false, justification=centering}
	\centering
	\subfloat[\label{fig:img_rod_s0} ]{\includegraphics[scale=0.465]{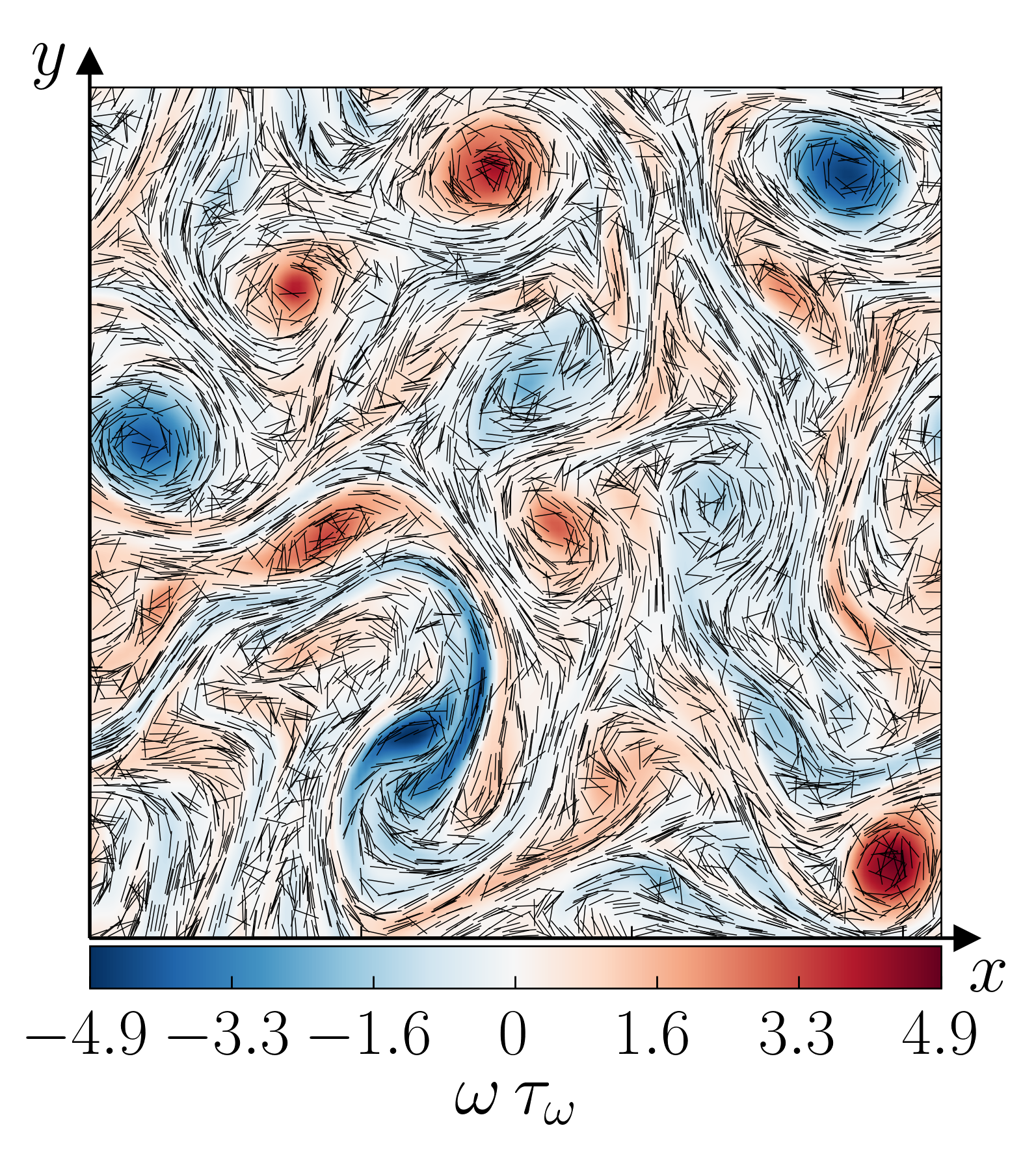}}
	\subfloat[\label{fig:img_rod_s2.6} ]{\includegraphics[scale=0.465]{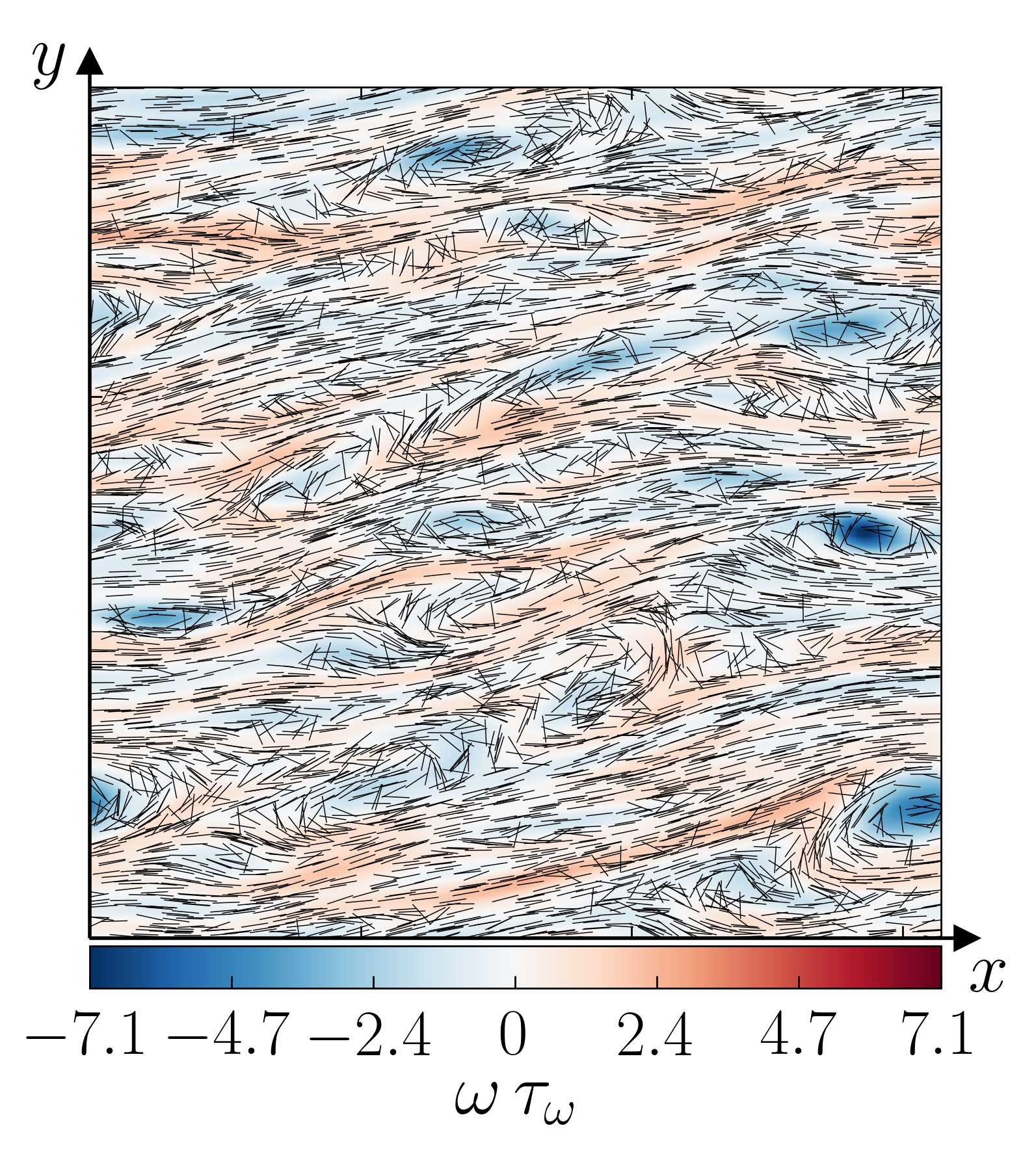}}
	\subfloat[\label{fig:PDF_theta} ]{\includegraphics[scale=0.45]{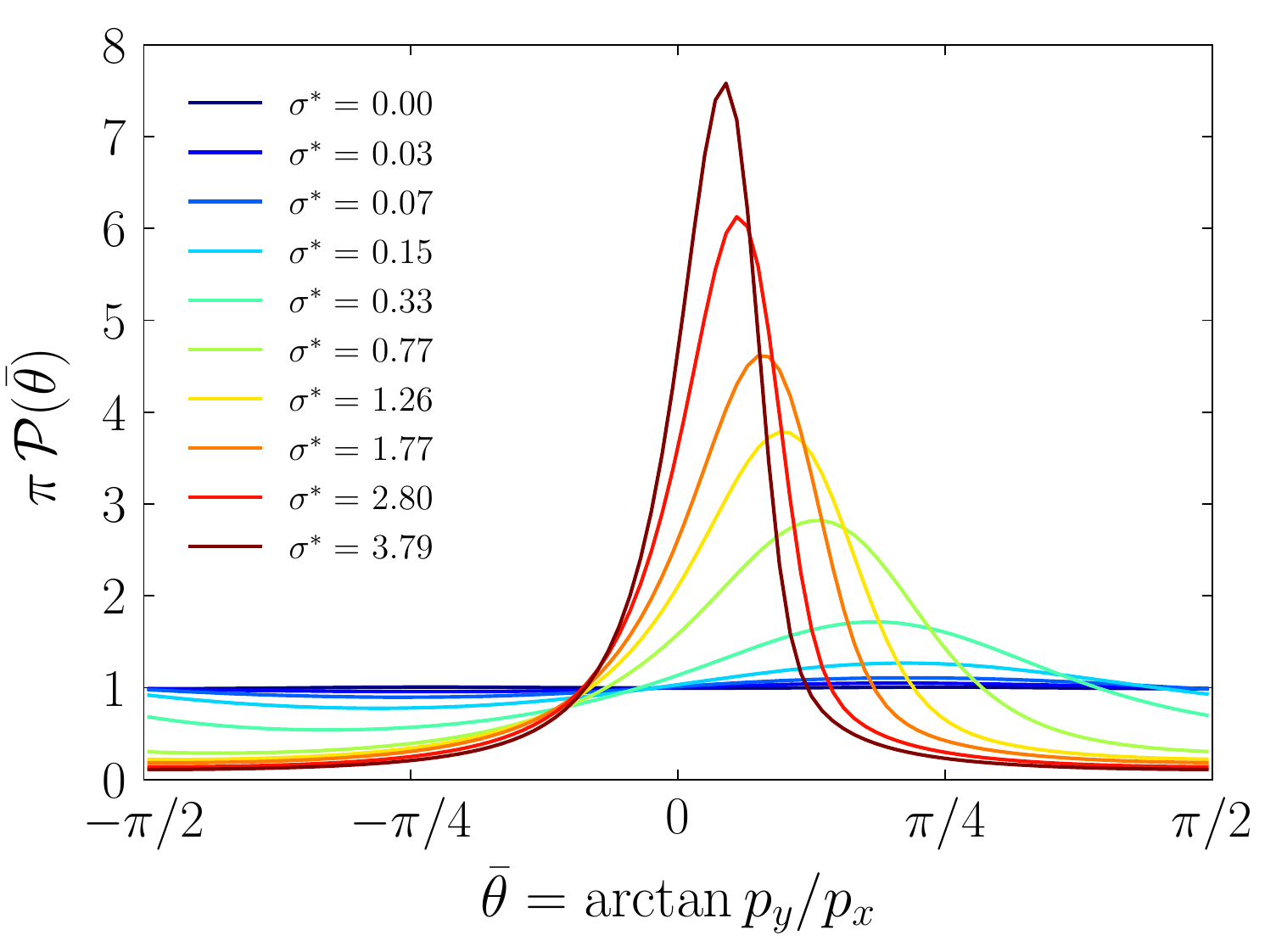}}
	\caption{\label{fig:snapshot_rods} (Colour online) 
		\protect\subref{fig:img_rod_s0},\protect\subref{fig:img_rod_s2.6} Vorticity field 
		$\omega$ of the turbulent fluctuation (coloured background) for two different values of the 
		shear rate along the $y$ direction for $\sigma^*=0$ and $\sigma^*=2.8$, respectively. Blue 
		corresponds to negative values (cyclonic eddies) and red to positive values (anticyclonic). On 
		the top of this field, the orientations of fibres are shown as black segments. 
		\Rep{\protect\subref{fig:PDF_theta}} Distribution of the rods angle $\bar{\theta}_t = \arctan 
		(p_y/p_x)$ with respect to the horizontal for various values of the shear.}
\end{figure}
In the presence of shear, rods display an anisotropic orientation and tend to preferentially align 
with the direction of the mean velocity gradient. This is qualitatively evidenced from 
Fig.~\ref{fig:snapshot_rods}\protect\subref{fig:img_rod_s0}-\protect\subref{fig:img_rod_s2.6},
 which show as black segments the instantaneous direction of rods without and with shear obtained 
 from DNS in the developed turbulent regime. In these figures, the underlying vorticity field is also 
 shown as a coloured background. One clearly observes a significant correlation between the 
 spatial distribution of rod orientations and the vorticity field. Indeed, rods tend in both cases to align 
 with the isoline of $\omega$ as already observed in the absence of shear~\cite{gupta2014}, and to 
 experience a strong rotation when captured by vortical structures. The presence of shear causes 
 vortical structures and vorticity filaments, \textit{i.e.}\ strips of constant vorticity, to become 
 anisotropic, with filaments stretched towards the shear direction and anticyclonic vortices (positive 
 vorticity) being fully depleted. This has a clear impact on the orientation of rods.

To quantify this preferential alignment, we report in Fig.~\ref{fig:snapshot_rods}\protect\subref{fig:PDF_theta} numerical 
measurements of the PDF of the rods folded angle $\bar{\theta}$ for various values of the shear 
parameter $\sigma^*$.  While the distributions are almost uniform when shear is weak, they develop 
an increasingly concentrated peak that move towards $\bar{\theta}=0$ when $\sigma^*$ increases. 
The distribution of orientation can be qualitatively understood in terms of the modifications of the 
vorticity distribution induced by shear. The concentration of the distribution towards 
$\bar{\theta}=0$ indeed results from the increasing alignment of vortex filaments with the horizontal. 
These results suggest that the distribution of $\bar{\theta}$ represents a direct measure of the flow 
anisotropies and can be used as a proxy to estimate the pitch angle of vorticity filaments with 
respect to the horizontal. However, estimating quantitatively such an alignment requires 
understanding how mean shear and turbulent fluctuations combine and compete in the orientation 
dynamics. Our aim is next to formulate a stochastic model for the time evolution of $\theta_t$ that 
accurately catches such effects.

\section{A stochastic model for rods orientation}\label{sec:orientation_model}

\subsection{Reproducing the separation between infinitesimally close fluid elements}
\label{subsec:evolmatrix}
We propose here a Lagrangian stochastic model for the orientation dynamics of rod-like particles, 
with the aim to properly balance the influences of mean shear and turbulent fluctuations and the 
intricate interplay between spatial and temporal properties of the velocity field.  The proposed 
approach relies on the relation between the orientation of a small, inertialess rod (infinitely thin 
spheroid) and the infinitesimal separation between two fluid trajectories~\cite{ni2014alignment}.  
The orientation $\bm p(t)$ that solves~\eqref{eq:jeffery_rod} is nothing but the normalised direction 
$\bm p = \bm r/|\bm r|$ of the separation vector that follows the tangent dynamics $\dd 
\bm r/\dd t = \mathbb{A}(t)\,\bm r$ along the reference tracer trajectory $\brmX(t)$. The linearity of 
this time evolution allows to express infinitesimal separations in terms of an evolution matrix: 
\begin{equation}\label{eq:solve_separation}
	\bm r(t) = \mathbb{D}(t,t_0)\,\bm r(t_0), \qquad \mbox{with } \quad \frac{\dd}{\dd t} \mathbb{D}(t,t_0) = \mathbb{A}(t)\, \mathbb{D}(t,t_0) \quad\mbox{and}\quad \mathbb{D}(t_0,t_0)=\mathbb{1}.
\end{equation}
The Lagrangian deformation tensor $\mathbb{D}(t,t_0)$ can be written in terms of a time-ordered 
matrix exponential
\begin{equation}\label{eq:evolution_matrix}
	\mathbb{D}(t,t_0) =
	\mathcal{T}\!\exp\left[\int_{t_0}^{t}  \mathbb{A}(s)\,\dd s \right] 
	\equiv \mathbb{1}
	+ \sum_{n=1}^{\infty} \int_{t_0}^{t}\int_{t_0}^{s_n}\!\!\!\cdots\!\!\int_{t_0}^{s_2} \mathbb{A}(s_n)\,\mathbb{A}(s_{n-1}) \cdots \mathbb{A}(s_1)\,\dd s_1 \cdots \dd s_{n-1}\, \dd s_n.
\end{equation}
It characterises the full distortion history between times $t_0$ and $t$ of infinitesimal fluid elements 
along the Lagrangian trajectory $\brmX(t)$. The singular values of $\mathbb{D}(t,t_0)$ describe 
how infinitesimal distances, areas, volumes grow under the action of fluid-velocity gradients and 
have been extensively studied to quantify Lagrangian chaos and turbulent mixing in the so-called 
Batchelor's regime~\cite{pierrehumbert1993global, balkovsky1999universal, borgas2004high, 
peacock2013lagrangian}.  However, when interested in the orientation statistics of small rods, 
relevant information is provided by the dynamics of angles and not that of amplitudes alone.

Our approach relies on the idea that an adequate model should reproduce the long-term effects of 
the deformation matrix~\eqref{eq:evolution_matrix}, rather than approximating directly the 
Lagrangian fluid-velocity gradient $\mathbb{A}$.  An important property of the evolution matrix is 
that it forms a semi-group and can thus be written as the ordered product of $N =  (t-t_0)/\Delta t$ 
intermediate evolution matrices over an arbitrary time step $\Delta t$: $\mathbb{D}(t, t_0) = 
\mathbb{D}(t,t_{N-1})\cdots\mathbb{D}(t_2,t_1)\,\mathbb{D}(t_1,t_0)$, where $t_n - t_{n-1} = \Delta 
t$. A proper statistical model for the intermediate matrices $\mathbb{D}(t_n,t_{n-1})$ ensures 
reproducing the long-term properties of $\mathbb{D}(t_0,t)$.  This decomposition can be 
interpreted as a time filtering that smooths out the detailed instantaneous fluctuations of the 
fluid-velocity gradient $\mathbb{A}$. The next step consists in adequately choosing the 
coarse-graining time $\Delta t$ with regard to the relevant timescales of the dynamics.

The fist choice consists in assuming that $\Delta t$ is much shorter than the inverse of the typical amplitude of $\mathbb{A}$, which is of the order of $\tau_\omega$ (see Sec.~\ref{subsec:shear2d}). This choice allows retaining only leading terms in the time-ordered exponential
\begin{equation}\label{eq:develop_evolution_matrix}
	\mathbb{D}(t_{n},t_{n-1}) = 
	\mathbb{1} + \langle\mathbb{A}\rangle\,\Delta t + \int_{t_{n-1}}^{t_n} \mathbb{A}'(s)\,\dd s +\smallO\! \left({\Delta t}/{\tauO}\right),
\end{equation} 
where we have introduced the fluctuating fluid-velocity gradient $\mathbb{A}' \equiv \mathbb{A} 
-\langle\mathbb{A}\rangle$.  Additionally, in order for such a model to be analytically tractable, we 
require that the intermediate evolution matrices be well approximated by independent Gaussian 
variables. This requires choosing $\Delta t$ much larger than the maximal integral correlation time of 
the Lagrangian velocity gradient $\tauI = \max{(\tauI^{ijkl})}$ with
\begin{equation} \label{eq:correls}
	\tauI^{ijkl} \equiv 
	\frac{\left|\mathcal{I}_{ijkl}\right|}{\C_{ijij}^{1/2}(0)\,\C_{klkl}^{1/2}(0)} \quad \mbox{with}\quad 
	\mathcal{I}_{ijkl} \equiv \frac{1}{2}\int_{-\infty}^{+\infty} \C_{ijkl}(\tau) \, \dd \tau \ \ \mbox{and}\ \ 
	\C_{ijkl}(\tau) = \braket{{\sf A}'_{ij}(0)\,{\sf A}'_{kl}(\tau)} \!.
\end{equation}
Here, summation is not assumed over repeated indices and we have used that Lagrangian velocity 
gradients are statistically stationary in time. Such a choice allows decomposing the integral in the 
right-hand side of~\eqref{eq:develop_evolution_matrix} as a sum of $M = \Delta t /\tauI\gg 1$ 
independent, identically distributed, random variables:
\[
	\int_{t_{n-1}}^{t_n} \mathbb{A}'(s)\,\dd s = \sum_{m=1}^M \int_{(m-1)\,\tau_{\rm I}}^{m\,\tau_{\rm 
	I}} \!\!\!\!\mathbb{A}'(t_n+s)\,\dd s \quad\mbox{with}\  \left\langle  \int_{(m-1)\,\tau_{\rm 
	I}}^{m\,\tau_{\rm I}} \!\!\!\!\mathsf{A}'_{ij}(t_n+s)\,\dd s  \int_{(m'-1)\,\tau_{\rm I}}^{m'\,\tau_{\rm I}} 
	\!\!\!\!\mathsf{A}'_{kl}(t_n+s)\,\dd s\right\rangle \approx 2\tauI\,\mathcal{I}_{ijkl}\,\delta_{m,m'}.
\]
Then, applying the central-limit theorem to this sum, one gets
\begin{equation}\label{eq:develop_evolution2}
	\mathbb{D}(t_{n},t_{n-1}) \stackrel{\rm law}{=}  \mathbb{1} + \langle\mathbb{A}\rangle\,\Delta t + \mathbb{S}^{(n)} +\smallO\!\left({\Delta t}/{\tauO}\right) + \smallO\!\left({\Delta t}/\tauI\right)^{1/2},
\end{equation} 
where the $\mathbb{S}^{(n)} $'s are Gaussian random matrices whose elements have mean and covariances
\begin{equation}
	\left\langle {\sf S}^{(n)}_{ij}\right\rangle = 0,\quad \left\langle {\sf S}^{(n)}_{ij}\,{\sf S}^{(n')}_{kl} \right\rangle = 2\Delta t\,\mathcal{D}_{ijkl} \,\delta_{n,n'} \mbox{ and }  \mathcal{D}_{ijkl} \approx \mathcal{I}_{ijkl}.
\end{equation}
Before extending further this approach, let us briefly comment on its validity. The proposed model relies on two assumptions: The first,  $\Delta t \ll \tauO$, allows to consider only the leading term of the time-ordered exponential~\eqref{eq:evolution_matrix}; The second, $\Delta t \gg \tauI$, consists in assuming that scales at which we observe the system are much longer than the correlation time of the fluid-velocity gradients, allowing to assume Gaussian statistics for not-too-large fluctuations. Finding out an adequate coarse-graining time scale $\Delta t$ such that $\tauI \ll \Delta t \ll \tauO$ requires a sufficient scale separation that is ensured by a small value of the \emph{Kubo number} $\Ku=\tauI/\tau_\omega \ll 1$.   This quantity is a dimensionless measurement of the Lagrangian correlation time with respect to the typical amplitude of velocity gradients.  In three-dimensional turbulent flow, this number in known to be at least of the order of one.  As we will later see (Sec.~\ref{03subsec:correlations}), the absence of vortex stretching and the presence of long-standing structures in two dimensions are responsible for observing rather large values of $\Ku$.  For that reason, we keep free how to select the covariance tensor $\mathcal{D}_{ijkl}$ in order to compare different alternatives. Nevertheless and in spite of the apparent inconsistency of the proposed approach, we will see in next section that the choice $\mathcal{D}_{ijkl} = \mathcal{I}_{ijkl}$ fairly reproduces low-order statistics for the orientation of rod particles.

The next step in our approach consists in finding out the evolution equation that is satisfied by a 
stochastic effective separation $\bm r^\star(t)$ whose statistics will reproduce those of the original 
infinitesimal separation $\bm r(t)$.  As it is often the case in stochastic modelling, there is an 
ambiguity on how to interpret the multiplicative noise that appears in the stochastic differential 
equation (SDE) followed by $\bm r^\star$.  The choice between \Ito and \Stra interpretations is here 
settled by the idea that the evolution matrix associated to the model dynamics should reproduce, to 
leading order, the right-hand side of Eq.~\eqref{eq:develop_evolution2}.  Using generalisations of 
the time-ordered exponential series \eqref{eq:evolution_matrix} to the case of 
SDEs~\cite{castell1993asymptotic}, we find that the appropriate interpretation is \Stra's. The 
evolution equation for $\bm r^\star(t)$ is then recovered by replacing 
Eq.~\eqref{eq:develop_evolution2} into Eq.~\eqref{eq:solve_separation}, leading to
\begin{equation}\label{eq:separation_model}
	\dd \bm r^\star =  \braket{\mathbb{A}} \, \bm r^\star \, \dd t
	+ (\mathbb{B}\, \partial \mathbb{W}_t) \bm r^\star.
\end{equation}
Here,  $\mathbb{B}$ is a deterministic tensor such that $\mathsf{B}_{ijmn}\mathsf{B}_{klmn} = 2 \, \mathcal{D}_{ijkl}$, the $d\times{d}$ random matrix $\mathbb{W}$ is made of independent Brownian motions, $d$ denoting the space dimension, and the differential $\partial$ emphasises that the stochastic integral has to be interpreted in \Stra's sense. A statistical model $\bm p^\star(t)$ for the orientation vector of rod particles in two-dimensional homogeneous shear flow is then obtained by prescribing $d=2$, that $\braket{\mathsf{A}_{ij}} = \sigma \delta_{i,1}\delta_{j,2}$ and by normalising the separation vector solution of Eq.~\eqref{eq:separation_model} by its length, \ie $\bm p^\star(t)=\bm r^\star(t)/|\bm r^\star(t)|$, as in the deterministic case. Note that the considerations leading to such a model are valid in any spatial dimension $d$.

Quantitative comparisons between the statistical model for the orientation $\bm p(t)$ and the results of DNS require properly matching timescales. In the above model, the relative importance of the average gradient compared to its fluctuations is measured through the ratio $\sigma/\|\mathcal{D}\|$, which balances the mean shear to a norm of the correlation tensor $\mathcal{D}_{ijkl}$ and thus, in principle, to the long-term Lagrangian statistics of fluid-velocity gradients.  However, we have seen in Sec.~\ref{subsec:shear2d} that the relevant way to non-dimensionalise the results of DNS is to introduce the turnover time $\tau_\omega$, defined as the inverse of the root-mean-squared vorticity, and thus to measure the relative importance of mean shear through the dimensionless parameter $\sigma^* = \tau_\omega\,\sigma$. Matching the model to DNS thus requires calibrating the correlation amplitude  $\|\mathcal{D}\|$ in units of $\tau_\omega$. This will be done by introducing an adjustable dimensionless multiplicative parameter.

\subsection{An effective stochastic model over angles}
\label{subsec:model}
We here turn back to the two-dimensional case to write, in a similar way to the deterministic dynamics~\eqref{eq:orientation}, a stochastic model equation for the evolution of the folded orientation angle $\btheta_t \equiv \arctan(p_y^\star(t)/p_x^\star(t)) = \arctan(r_y^\star(t)/r_x^\star(t))$ with values in $[-\nicepi, \nicepi]$. Note that, in the sake of simplifying notations, we dropped here the $^\star$  symbol used to refer to statistically modelled quantities.  Applying the \Ito formula to Eq.~\eqref{eq:separation_model} (see Appendix~\ref{apndx:2d_ito_lemma}),  $\bar{\theta}_t$ is found to follow
\begin{equation}\label{eq:folded_theta}
\btheta_t = \text{cmod} \left(\btheta_0 + \int_0^t a(\btheta_s) \,\dd s + \int_0^t b(\btheta_s) \,\partial W_s\right),
\end{equation}
where $W_t$ is a one-dimensional Brownian motion, $\text{cmod}(x) := ((x + \nicepi)\, \text{mod }  \pi) - \nicepi$,   and the stochastic integral is again understood with the Stratonovich convention. The drift and diffusion coefficients are respectively,
\begin{equation}
	a(\btheta) =
	\frac{\s^*}{2}\left(\cos(2\btheta)-1\right), \quad
	b(\btheta) =
	\left(\gamma_0 +\gamma_1\sin(2\btheta) 
	+\gamma_2\sin(4\btheta)
	+\gamma_3 \cos(2\btheta) 
	+\gamma_4 \cos(4\btheta)\right)^{1/2}. \label{eq:coeff_sde}
\end{equation}
The $\gamma_n$ parameters are expressed in terms of the effective diffusion tensor $\D_{ijkl}$ 
appearing in \eqref{eq:separation_model} through
\begin{gather} \label{eq:defgammas}
\begin{gathered}
\begin{array}{lcl}
		\gamma_0 = \D_{1111} + \tfrac{1}{2}\,\D_{1221} +\tfrac{3}{4}\,(\D_{2121}+\D_{1212}),&\strut\qquad\strut&
		\gamma_1 = 2\,(\D_{1112} - \D_{1121}), \\
		\gamma_2 =  -\D_{1112} - \D_{1121},&&
		\gamma_3 = \D_{2121} - \D_{1212},
	\end{array}	\\
	\gamma_4 = -\D_{1111} + \tfrac{1}{2}\,\D_{1221} + \tfrac{1}{4}\,(\D_{2121} + \D_{1212}).
\end{gathered}
\end{gather}

It is important to notice that the model \eqref{eq:folded_theta}-\eqref{eq:coeff_sde} for orientation dynamics involves diffusive terms that are proportional to $\cos(4\btheta)$ and $\sin(4\btheta)$, in contrast with its deterministic, smooth counterpart of Sec.~\ref{subsec:rods_dynamics}. This is a consequence of the projection of the diffusive dynamics of $\bm r^\star(t)$ onto the unit sphere and such terms would not have been present if the model had been directly derived from the physical angle dynamics~\eqref{eq:orientation}. The considerations of previous section however show that our approach is a consistent manner to consider the cumulative effects of the fluid-velocity gradient.

In view of comparing the statistical model to DNS, we first focus on the stationary distribution of rod orientations. The probability density function $\mathcal{P}(\btheta, t)$ of the stochastic folded angle satisfies the \FP equation
\begin{equation} \label{eq:FP_continuity}
	\dt \mathcal{P}(\btheta, t) 
	+ \partial_{\btheta} j(\btheta, t) = 0,
\end{equation}
where the \emph{probability current}, which represents the amount of probability flux across a given point $\btheta$, reads
\begin{equation} \label{eq:Flux_continuity}
	j(\btheta, t) \equiv a(\btheta)\,\mathcal{P}(\btheta, t) 
	-\frac{1}{2} b(\btheta)\,\partial_{\btheta}
	\left[ b(\btheta)\,\mathcal{P}(\btheta, t)\right].
\end{equation}
The \FP equation \eqref{eq:FP_continuity} is supplemented with periodic boundary conditions $\Pdf(-\nicepi, t) = \Pdf(\nicepi, t)$, and the usual normalisation of the probability density function $\int_{-\nicepi}^{\nicepi}\Pdf(\btheta, t) \,\dd\btheta = 1$. 

It has been shown in~\cite{bensoussan2011asymptotic} that the process $\btheta_t$ is an 
exponentially ergodic diffusion, ensuring that its distribution reaches exponentially fast at long times 
a uniquely defined statistical steady state given by the stationary solution $\Pst(\btheta)$ of 
\eqref{eq:FP_continuity}. This solution corresponds to a constant probability current, $j(\btheta, t) 
= -\mathcal{J}$, and satisfies
\begin{align}\label{eq:FP_stationary}
	\begin{aligned}
		\partial_{\btheta}\left[b(\btheta)\,\Pst(\btheta)\right] = \frac{2a(\btheta)}{b(\btheta)}
		 \Pst(\btheta) +\frac{2 \mathcal{J}}{b(\btheta)}.
	\end{aligned}
\end{align}
It is important to notice that such a solution exists only if the diffusion coefficient $b(\btheta)$ never vanishes. This has been verified numerically for the various values of the $\gamma_n$'s parameters used in this study (see Appendix \ref{apndx:strict_pos_b}). Integrating \eqref{eq:FP_stationary} leads to
\begin{eqnarray}
	&& \Pst(\btheta)= 
	\frac{{\rm e}^{\Psi(\btheta)}}{\mathcal{N}\,b(\btheta)}
	\left(1 +2\mathcal{J}\mathcal{N}\int_{-\nicepi}^{\btheta} \frac{{\rm e}^{-\Psi(\btheta')}}{b(\btheta')} \dd\btheta'\right),
	\qquad\mbox{with} \quad  \Psi(\btheta)= \int_{-\nicepi}^{\btheta} \frac{2 a(\btheta')}{b^2(\btheta')} \dd \btheta',\label{eq:PDF_stationary}\\
&& \mbox{and} \quad
	\mathcal{N} = \left[\int_{-\nicepi}^{\nicepi} \frac{{\rm e}^{\Psi(\btheta)}}{b(\btheta)}\dd\btheta\right] \left[ 1-2\mathcal{J} \int_{-\nicepi}^{\nicepi} \frac{{\rm e}^{\Psi(\btheta)}}{b(\btheta)} \int_{-\nicepi}^{\btheta} \frac{{\rm e}^{-\Psi(\btheta')}}{b(\btheta')} \dd\btheta'\,\dd\btheta\right]^{-1}.
\end{eqnarray}
The constant $\mathcal{N}$ ensures here the normalisation of $\Pst$.  The probability current $\mathcal{J}$ is obtained by imposing periodic boundary conditions: $\Pst(\nicepi) = \Pst(-\nicepi)$.  This leads to
\begin{equation}
	\mathcal{J} = \frac{1}{2} \left[ \frac{1}{{\rm e}^{-\Psi(\nicepi)}-1}\int_{-\nicepi}^{\nicepi} \frac{{\rm e}^{\Psi(\btheta)}}{b(\btheta)} \dd\btheta \int_{-\nicepi}^{\nicepi} \frac{{\rm e}^{-\Psi(\btheta)}}{b(\btheta)} \dd\btheta+\int_{-\nicepi}^{\nicepi} \frac{{\rm e}^{\Psi(\btheta)}}{b(\btheta)} \int_{-\nicepi}^{\btheta} \frac{{\rm e}^{-\Psi(\btheta')}}{b(\btheta')} \dd \btheta'\,\dd\btheta\right]^{-1}.
\end{equation}

\subsection{Comparison of different models for the correlation tensor}\label{03subsec:correlations}

To contrast the above analytical stationary distribution with the result of DNS, we need to prescribe 
the effective correlation tensor $\D_{ijkl}$ that enters the definitions 
\eqref{eq:coeff_sde}-\eqref{eq:defgammas} of the model diffusion coefficient $b$. We consider and 
compare three different choices, with an increasing level of complexity.

The first case, which is the most straightforward, consists in assuming that fluctuations are isotropic. Together with the incompressibility constraint, this leads to write the effective correlation tensor as
\begin{equation}
	\D_{ijkl} \approx \Diso_{ijkl} \equiv
	\frac{ \aliso }{\tauO} \,
	(3\delta_{ik}\delta_{jl} -\delta_{ij}\delta_{kl} -\delta_{il}\delta_{jk}).
\end{equation}
This approximation follows most common approaches in statistical models of Lagrangian turbulence. The parameter $\aliso$ is often interpreted as a Kubo number and links the instantaneous properties of the flow (entailed in $\tauO$) to the long-term effect of gradients that the noise with correlations $\D_{ijkl}$ is expected to reproduce. In that case, noise correlations are independent of the mean shear $\sigma^*$. Also, when using such a form, the stochastic model of previous subsection is drastically simplified: all $\gamma_n$'s except  $\gamma_0$ vanish in the diffusion coefficient~\eqref{eq:coeff_sde}.

\begin{figure}[ht!]
\captionsetup[subfigure]{position=bottom, labelfont=bf,textfont=normalfont,	singlelinecheck=false,  justification=centering}
\centering
\subfloat[\label{fig:Cijkl0_sigma}]{
	\includegraphics[scale=0.48]{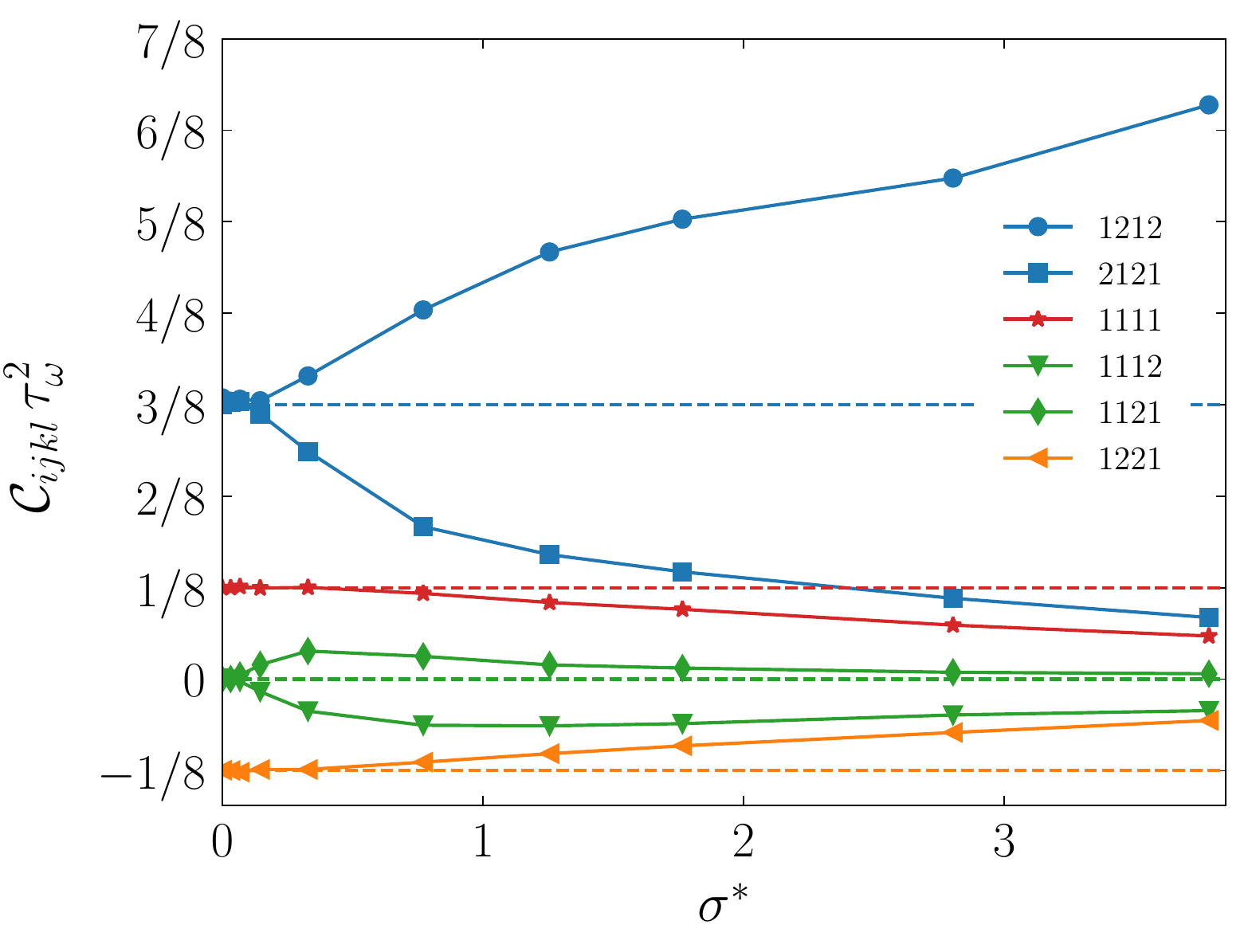}}
	\hfil
\subfloat[\label{fig:CijklE_sigma}]{\includegraphics[scale=0.48]{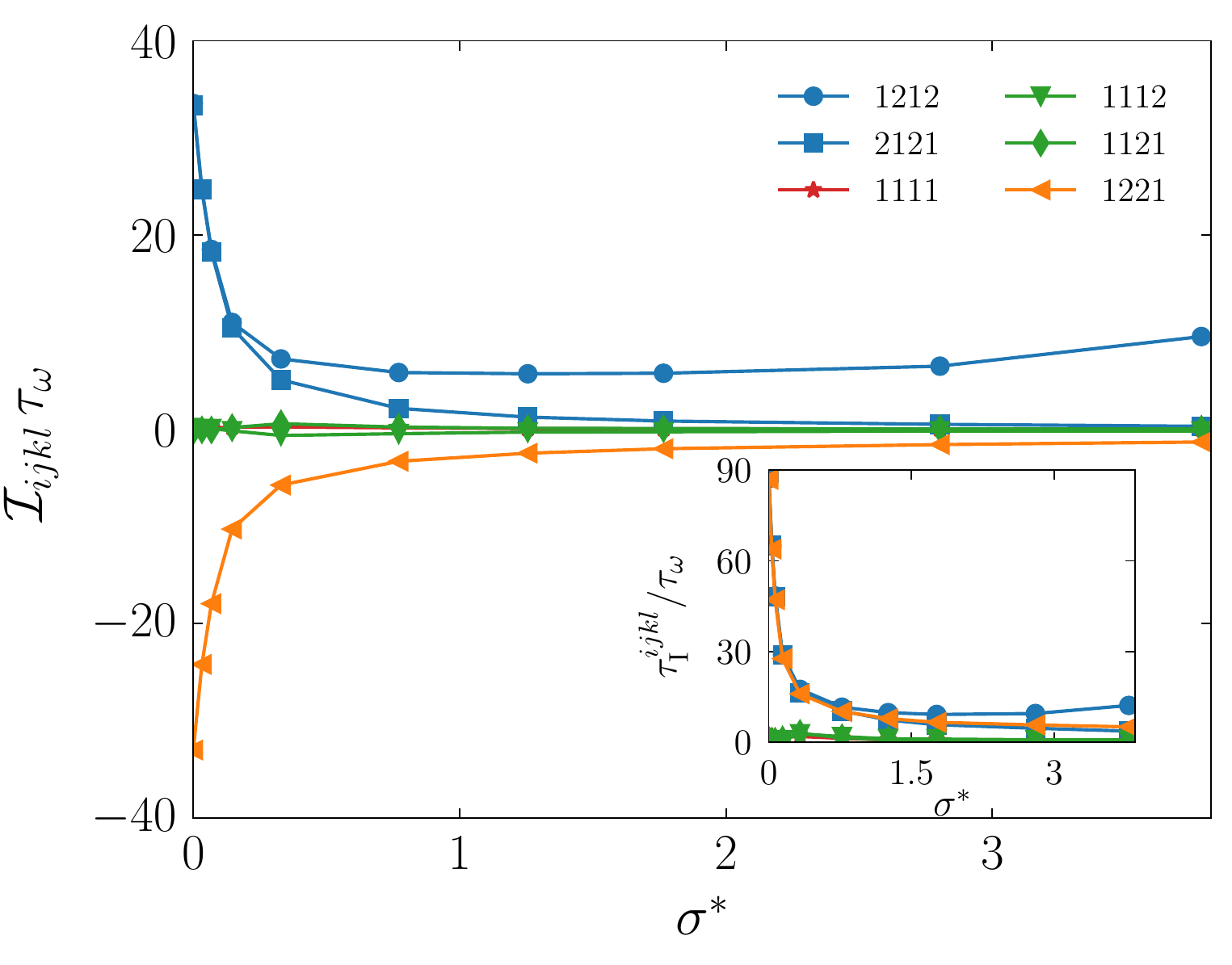}}
	\caption{\label{fig:sigma_correlations}
		(Colour online) \protect\subref{fig:Cijkl0_sigma} One-point, one-time correlation $\C_{ijkl}(0)$ of the fluid-velocity gradient tensor obtained from DNS, as a function of the dimensionless shear parameter $\sigma^* = \tauO\,\sigma$, where $\tauO = \langle \omega^2\rangle^{-1/2}$ . The horizontal dashed lines correspond to isotropic statistics. 
		\protect\subref{fig:CijklE_sigma} Integral correlation tensor $\mathcal{I}_{ijkl}$ measured as in 
		Eq.~\protect\eqref{eq:correls} as a function of $\sigma^*$. The inset shows the corresponding 
		component-dependent Lagrangian integral correlation time $\tauI^{ijkl}$.}
\end{figure}

The second case that we investigate consists in accounting for the single-time anisotropies of the fluid-velocity gradients that are caused by the mean shear.  The effective correlation tensor then reads
\begin{equation}
	\D_{ijkl} \approx \Daniso_{ijkl} \equiv \alaniso \, \tauO \, \C_{ijkl}(0),
\end{equation}
where, as previously, $\alaniso$ is a calibration parameter that can be interpreted as a Kubo number. 
$\C_{ijkl}(0)$ is here the equal-time covariance matrix of the fluid-velocity gradients, as defined in 
Eq.~\eqref{eq:correls}.  We then rely on DNS data to evaluate how this correlation tensor depends 
on the mean shear.  The six independent components of $\C_{ijkl}(0)$  are shown in 
Fig.~\ref{fig:sigma_correlations}\protect\subref{fig:Cijkl0_sigma} as a function of $\sigma^*$. Most important deviations from 
isotropy occur both for the component along the mean shear $\C_{1212}(0) = \braket{(\partial_y 
u_x)^2}$, which  increases as a function of $\s^*$, and for the transverse component $\C_{2121}(0) 
= \braket{(\partial_x u_y)^2}$, which is depleted by shear.

Finally, the third case follows the considerations of Sec.~\ref{subsec:evolmatrix} in order to better approximate the cumulative effects of the velocity gradient along Lagrangian paths. The noise correlations here reads 
\begin{equation} \label{eq:dintapprox}
	\D_{ijkl} \approx \Dint_{ijkl} \equiv \alint \, \mathcal{I}_{ijkl},
\end{equation}
where $\mathcal{I}_{ijkl}$ denotes the time-integral of the two-time Lagrangian covariance of the 
velocity gradient tensor and was defined in Eq.~\eqref{eq:correls}. In the same spirit as
before, a tuning parameter $\alint$ is introduced. This time, however, the dependence on correlation 
times is entailed in the definition of $\mathcal{I}_{ijkl}$, so that $\alint$ cannot be straightforwardly 
assimilated to a Kubo number. As discussed in Sec.~\ref{subsec:evolmatrix} it rather parametrises 
the relationship between single-time and averages fluctuations and depends, in principle, on the 
details of the underlying coarse-graining procedure. The six independent components of the integral 
tensor $\mathcal{I}_{ijkl}$ measured from DNS are shown in Fig.~\ref{fig:sigma_correlations}\protect\subref{fig:CijklE_sigma}. 
At $\s^*=0$, transverse components are markedly large, which is a very strong signature of the 
presence of long-living rotating structures in the flow. One observes there that 
$\mathcal{I}_{1212}=\mathcal{I}_{2121}=-\mathcal{I}_{1221}$, confirming the strongly circular 
Beltrami nature of the vortices. The measured values indicate that tracers typically spend tens of 
turnover times in such structures.  The effect of increasing $\sigma^*$ is to rapidly reduce such 
correlations, a signature of the depletion of structures by shear (see Sec.~\ref{subsec:shear2d}).  At 
larger values $\s^*\gtrsim 0.15$,  anisotropies start to develop, in particular for the component 
$\mathcal{I}_{1212}$ along the mean shear direction. Clearly, the behaviour of the integral 
correlations $\mathcal{I}_{ijkl}$ is very different from the single-time correlations $\C_{ijkl}(0)$ 
shown on Fig.~\ref{fig:sigma_correlations}\protect\subref{fig:Cijkl0_sigma}. Obviously, the two correlations are not related by a simple 
proportionality law. To assess the peculiar behaviour of $\Ku$ in two dimensional turbulence, the 
corresponding component-dependent are displayed in inset of Fig.~\ref{fig:sigma_correlations}\protect\subref{fig:CijklE_sigma}.

\begin{figure}[ht!]
\captionsetup[subfigure]{position=bottom, labelfont=bf,textfont=normalfont,	singlelinecheck=false,  justification=centering}
\centering
\subfloat[\label{fig:Mean_ThetaMDL_DNS}]{
\includegraphics[scale=0.48]{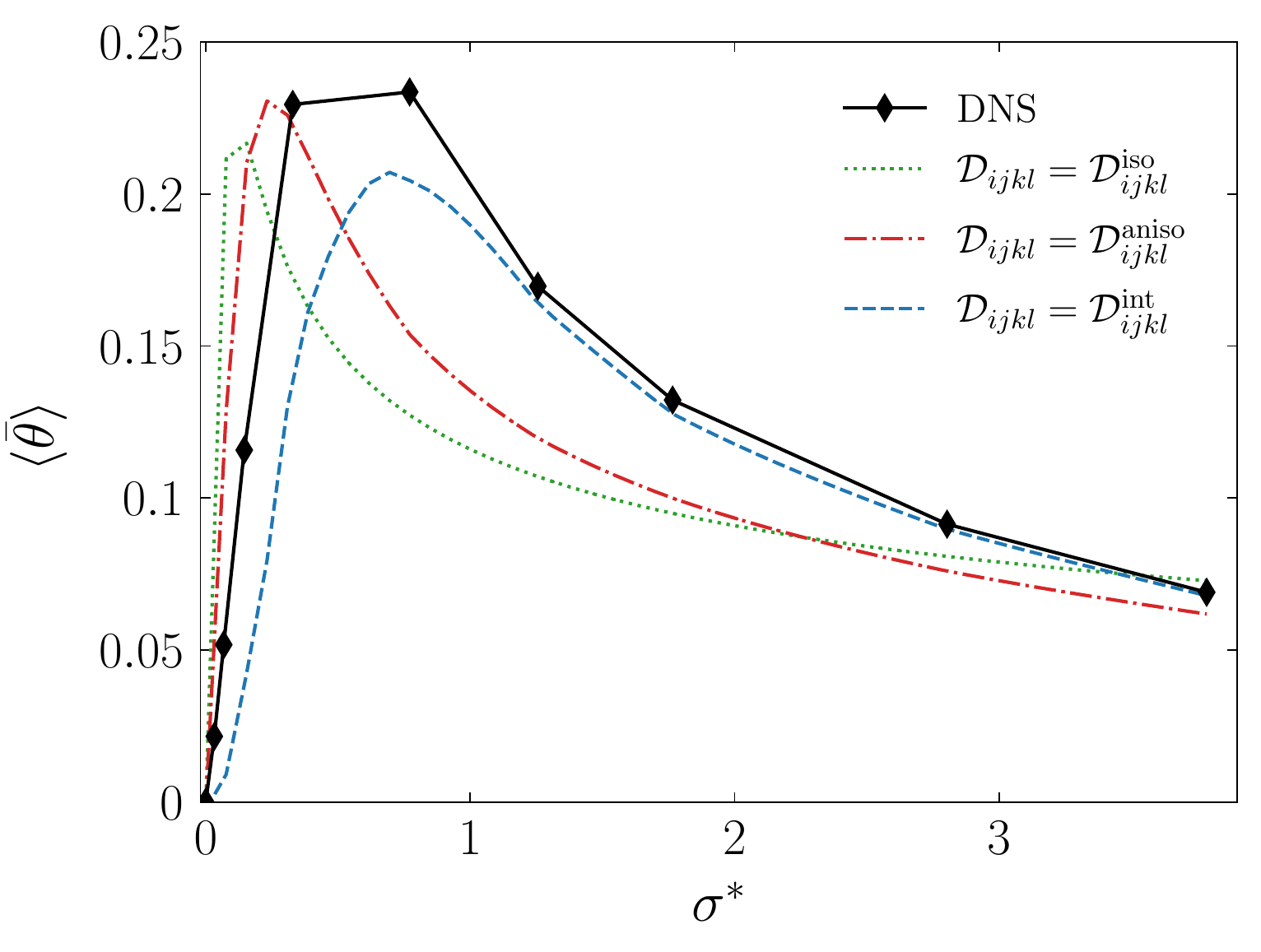}}
	\hfil
\subfloat[\label{fig:PDF_ThetaMDL3_DNS}]{
\includegraphics[scale=0.48]{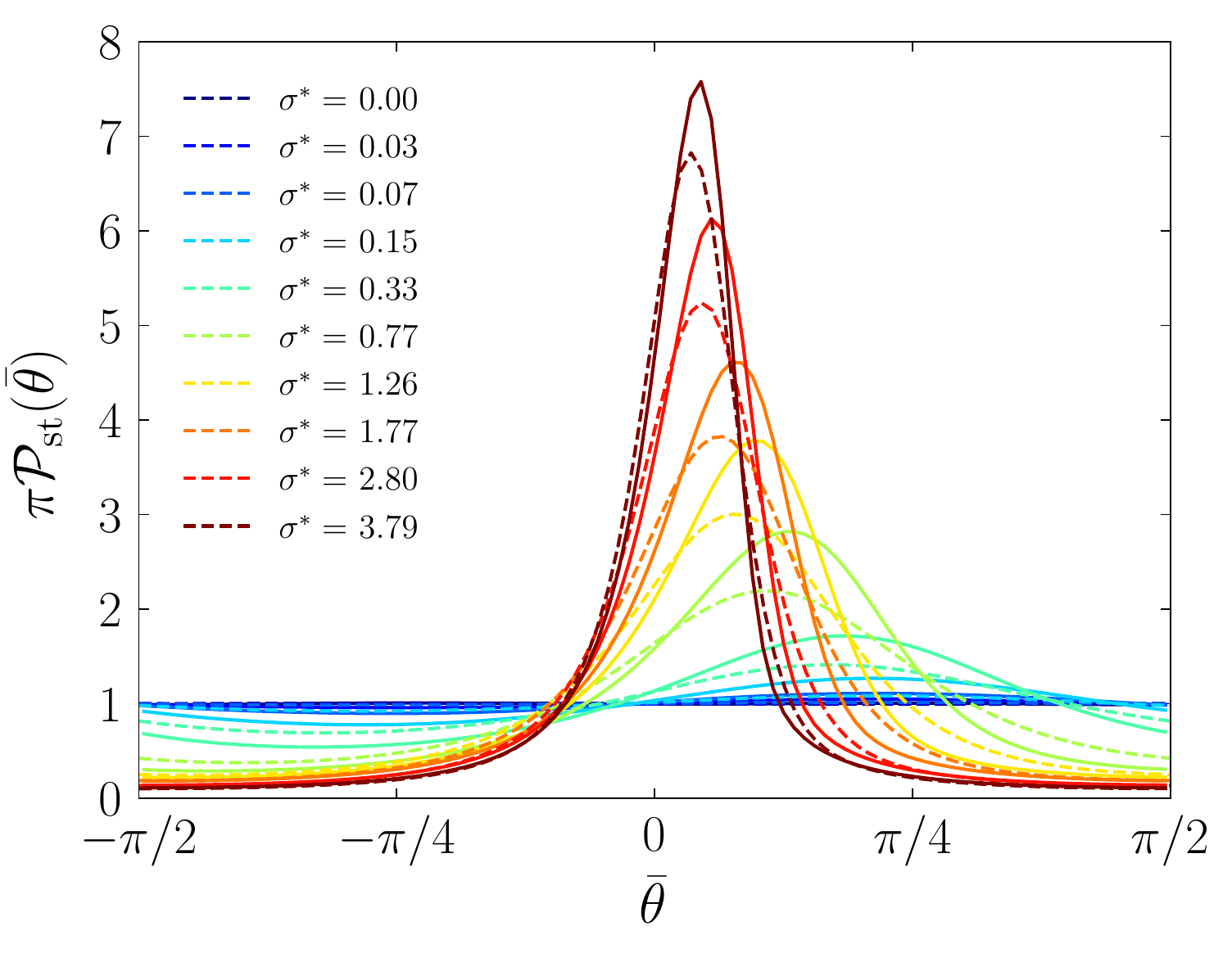}}
\caption{\label{fig:PDF_theta_mdl}
	(Colour online) \protect\subref{fig:Mean_ThetaMDL_DNS} Variation as a function of $\s^*$ of the 
	average orientation angle $\langle\btheta\rangle$ obtained from the DNS and the three models, as 
	labelled. The three selected parameters were $\aliso=0.076$, $\alaniso=0.189$, and 
	$\alint=0.053$.
	\protect\subref{fig:PDF_ThetaMDL3_DNS} Stationary distribution $\Pst$ of the rods folded angle 
	obtained with the integral correlation tensor $\Dint_{ijkl}$ with $\alint=0.053$ and for various 
	values of the shear (dashed lines). DNS measurements for different values of $\s^*$ are shown as 
	continuous lines.}
\end{figure}
The results of the three models are balanced and contrasted to DNS data. To make this tractable, we 
focus on the stationary distribution $\pst$ of the folded orientation angle and on its mean value 
$\langle\btheta\rangle = \int \btheta\,\pst(\btheta)\,\dd\btheta$. Different values of the free 
parameters $\aliso$,  $\alaniso$, and $\alint$ have been tested to minimise errors, not only in terms 
of $\braket{\btheta}$, but also for the peak of the stationary orientation distribution.

The results obtained with the isotropic correlation $\Diso_{ijkl}$ reproduce the DNS, only for small 
values of $\s^*$. When the shear rate increases, the stationary distribution becomes less peaked 
than the DNS (not shown), even for the optimal value $\aliso=0.076$. This behaviour is confirmed in 
Fig.~\ref{fig:PDF_theta_mdl}\protect\subref{fig:Mean_ThetaMDL_DNS}, which represents 
$\langle\btheta\rangle$ as a function of the shear rate. Discrepancies are important for all finite 
values of $\sigma^*$. The isotropic model is thus found to have a limited effectiveness to 
reproduce DNS statistics, even at a qualitative level.

When using the anisotropic tensor $\Daniso_{ijkl}$, the stationary distribution $\pst$ is in better agreement with DNS. The average value $\btheta$ shown in Fig.~\ref{fig:PDF_theta_mdl}\protect\subref{fig:Mean_ThetaMDL_DNS} displays clear improvements with respect to the isotropic model, in particular for large values of the shear rate. Still, this anisotropic model based on single-time gradients statistics reproduces only the trends that the presence of shear produces in the DNS. Its effectiveness is restricted to qualitative aspects.

Quantitative agreements between the model and the simulations are obtained when using the integral correlation tensor $\Dint_{ijkl}$.  This is clear for the average angle shown in Fig.~\ref{fig:PDF_theta_mdl}\protect\subref{fig:Mean_ThetaMDL_DNS}, in particular at large values of the shear parameter. Discrepancies can only be appreciated at intermediate values of $\s^*$, for which the model angular distribution is noticeably less peaked than in DNS. This can be seen in Fig.~\ref{fig:PDF_theta_mdl}\protect\subref{fig:PDF_ThetaMDL3_DNS}, which shows the stationary distribution $\pst$ of the rods angle both for the model with integral correlations and for DNS data. The agreement is the greatest at the largest values of $\s^*$.  This can be qualitatively explained. As discussed in Sec.~\ref{subsec:evolmatrix}, modelling the fluid-velocity gradient in terms of a white-noise with correlations $\Dint_{ijkl}$ requires a small Kubo number, and thus typical correlation times shorter than the turnover time $\tauO$. A strong shear is responsible for a shortening of the living time of coherent structures and thus of a decrease of correlation times that becomes of the same order as the shear timescale $\sigma^{-1}$ (see the inset of Fig.~\ref{fig:sigma_correlations}\protect\subref{fig:CijklE_sigma}). We thus expect that $\Ku <  1/\s^*$, so that the white-noise limit could be asymptotically reached in the limit of strong shear.

Above results on single-time stationary statistics for the orientation of rods suggest that the model~\eqref{eq:dintapprox} based on integral correlations gives the best approximation of velocity gradient fluctuations.  We hereafter concentrate on it and we next investigate its effectiveness for two-time statistics.

\section{Tumbling statistics}\label{sec:tumbling}
In situations where the average velocity gradient vanishes, the angular dynamics of non-spherical 
particles is usually measured in terms of the \textit{tumbling rate}, which is defined as the 
root-mean-squared rate of change $\langle|\dd\bm p/\dd t|^2\rangle^{1/2}$ of the orientation vector. 
Much numerical and experimental work has focused on tumbling rate statistics 
(see~\cite{voth2017anisotropic} for a review).  However, in the framework of stochastic Lagrangian 
models, the orientation vector $\bm p$ diffuses, preventing from properly defining the tumbling rate. 
Nevertheless, alternative quantities can be introduced in order to gain access to fluctuations of the 
rods rotation rates and to provide quantitative predictions.

Our idea consists in unfolding the orientation angle $\btheta_t\in[-\nicepi,\nicepi]$ to the whole real line, and to investigate the long-term evolution of the rods orientation in terms of the (non-stationary) dynamics of the unfolded angle $\theta_t\in\mathbb{R}$. This evolution is entailed in the angular displacement $\delta \theta_t = \theta_t -\theta_0$ that represents the change in time of the unfolded angle with respect to its initial position. In its classical definition, the tumbling rate describes the instantaneous rate of change of the orientation and enters the short-time evolution of the angular displacement: A Taylor expansion indeed gives  $\langle \delta \theta_t^2 \rangle \approx t^2 \langle|\dd\bm p/\dd t|^2\rangle$. At extremely long times, the angular displacement is expected to reach a diffusive regime where the variance of  $\delta \theta_t$ grows linearly with time. The stochastic model for rods orientation that was introduced in previous section should capture this ultimate behaviour.

\begin{figure}[ht!]
	\centering
	\includegraphics[scale=0.48]{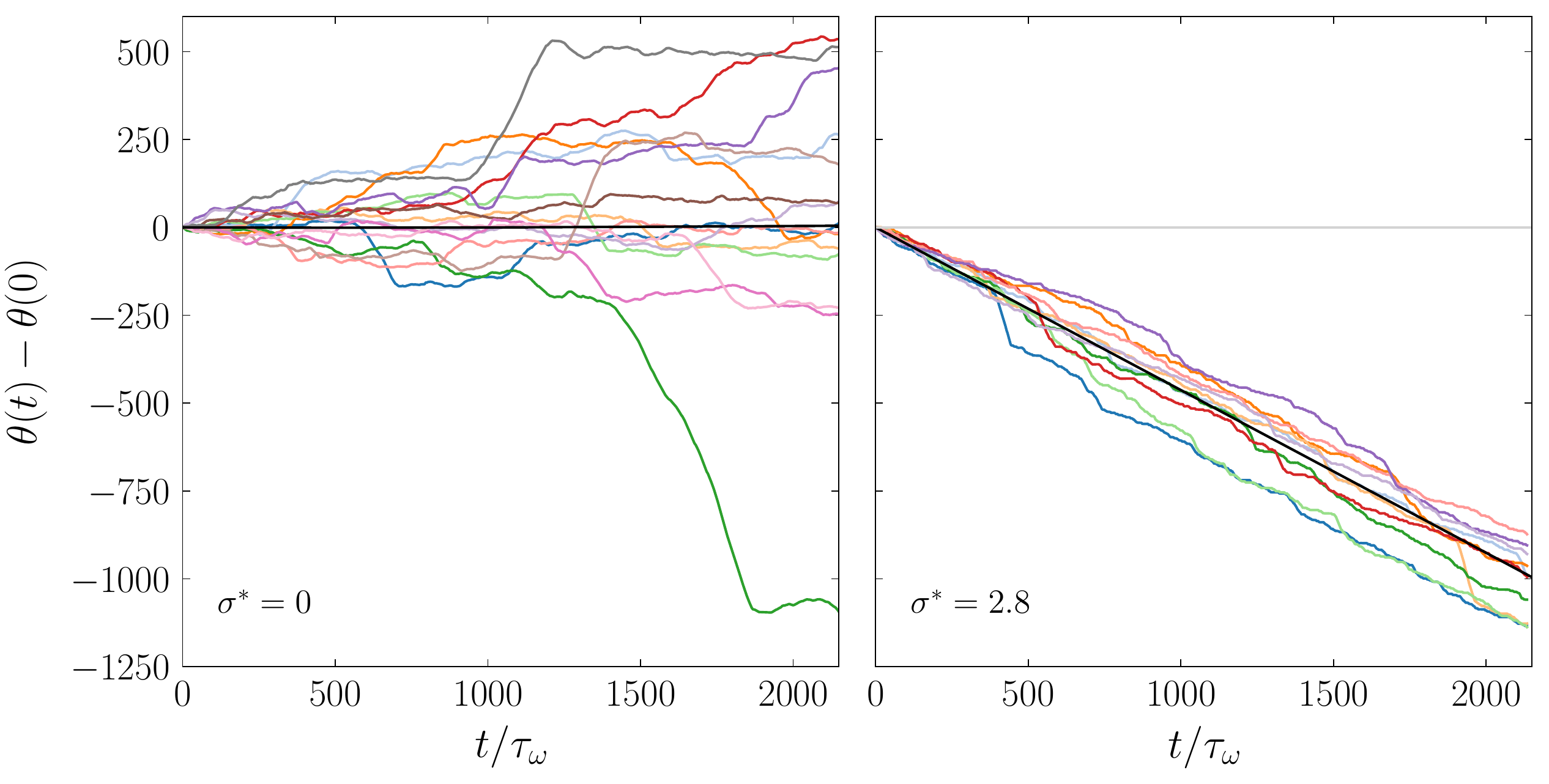}
	\caption{\label{fig:traj_Dtheta} 
		(Colour online) Typical trajectories of the angular displacement $\delta 
		\theta_t=\theta_t-\theta_0$  obtained  from DNS in the absence of shear (left panel) and for 
		$\sigma^*=2.8$ (right panel). Their average behaviour is shown as black lines.}
\end{figure}
Figure~\ref{fig:traj_Dtheta} shows several trajectories of the angular displacement extracted from 
DNS. In the absence of shear  $\sigma^*=0$ (left-hand panel), rod orientations fluctuate along their 
mean $\braket{\theta_t}=\theta_0$. Still, one qualitatively observes that at intermediate times, 
angular fluctuations can probably not be described in terms of a simple diffusion, because they 
involve strong quasi-ballistic excursions during which $\delta\theta_t$ varies linearly. This 
corresponds to events during which Lagrangian tracers are captured by vortex structures and, as we 
will see later, these long-range excursions have strong signatures in the distribution of angular 
increments.  In the 
presence of a mean shear $\sigma^*>0$ (right-hand panel), the average angular displacement 
$\langle\delta\theta_t\rangle$ decreases linearly as a function of time. This is a signature of the 
out-of-equilibrium dynamics reached by the folded orientation angle. Indeed, when 
$\sigma^*\to\infty$, the rod orientation has a fixed point at $\btheta = 0$ (\textit{i.e.}\  $\theta = 
0\;\mbox{mod}\;\pi$), where the local dynamics follows $\dd\btheta/\dd t \approx 
-\sigma^*\btheta^2$. This fixed point attracts in a finite time all trajectories approaching it from 
$\btheta>0$, but any perturbation that pushes the orientation angle to $\btheta<0$ triggers an 
instability and the rod initiates a Jeffery orbit. At finite values of $\sigma^*$, the fluctuations of 
velocity gradients are perturbing this quasi-equilibrium causing the rod to frequently tumble. In the 
stochastic model of previous section, this average angular displacement is entailed in the constant 
negative probability current $-\mathcal{J}$ that characterises the stationary distribution of the 
folded angle $\btheta_t$. DNS results moreover show that ballistic excursions of $\delta\theta_t$ 
are also present in the case with shear, but this time with a negative bias reflecting the skewness of 
the vorticity distribution.

As we will now see, such considerations lead to introduce two different measures for the rod tumbling rate: one reflecting the effect of a mean shear and obtained from the average angular displacement $\delta\theta_t$; the other associated to the long-time diffusive behaviour of $\delta\theta_t$ and obtained from its variance.

\subsection{Average angular displacement}\label{03subsec:mean_tumbling_rate}
\begin{figure}[h!]
\captionsetup[subfigure]{position=bottom, labelfont=bf,textfont=normalfont,	singlelinecheck=false,  justification=centering}
\centering
\subfloat[\label{fig:Mean_DTheta}]{
\includegraphics[scale=0.48]{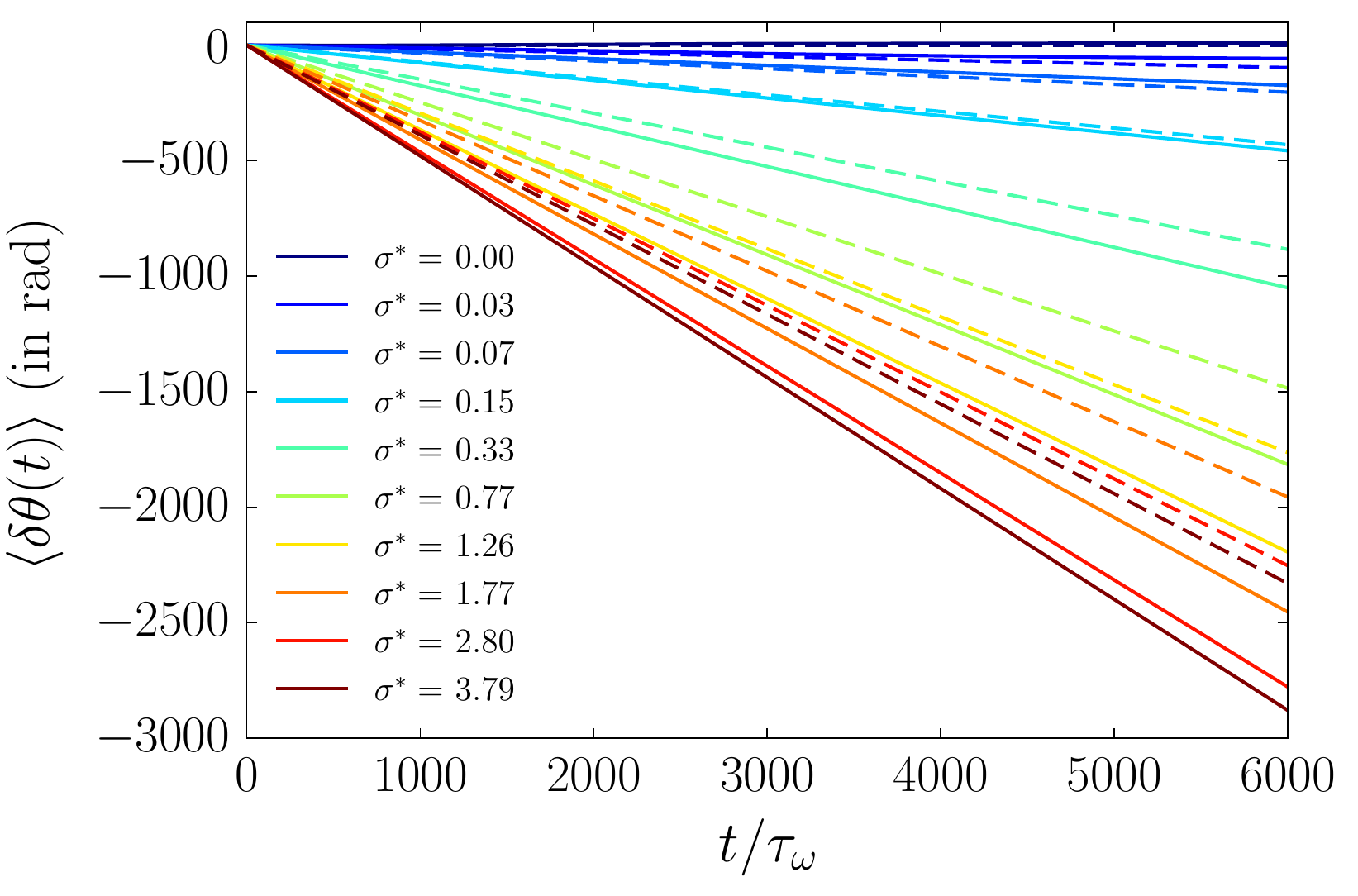}}
	\hfil
\subfloat[\label{fig:DThetaInf}]{
\includegraphics[scale=0.48]{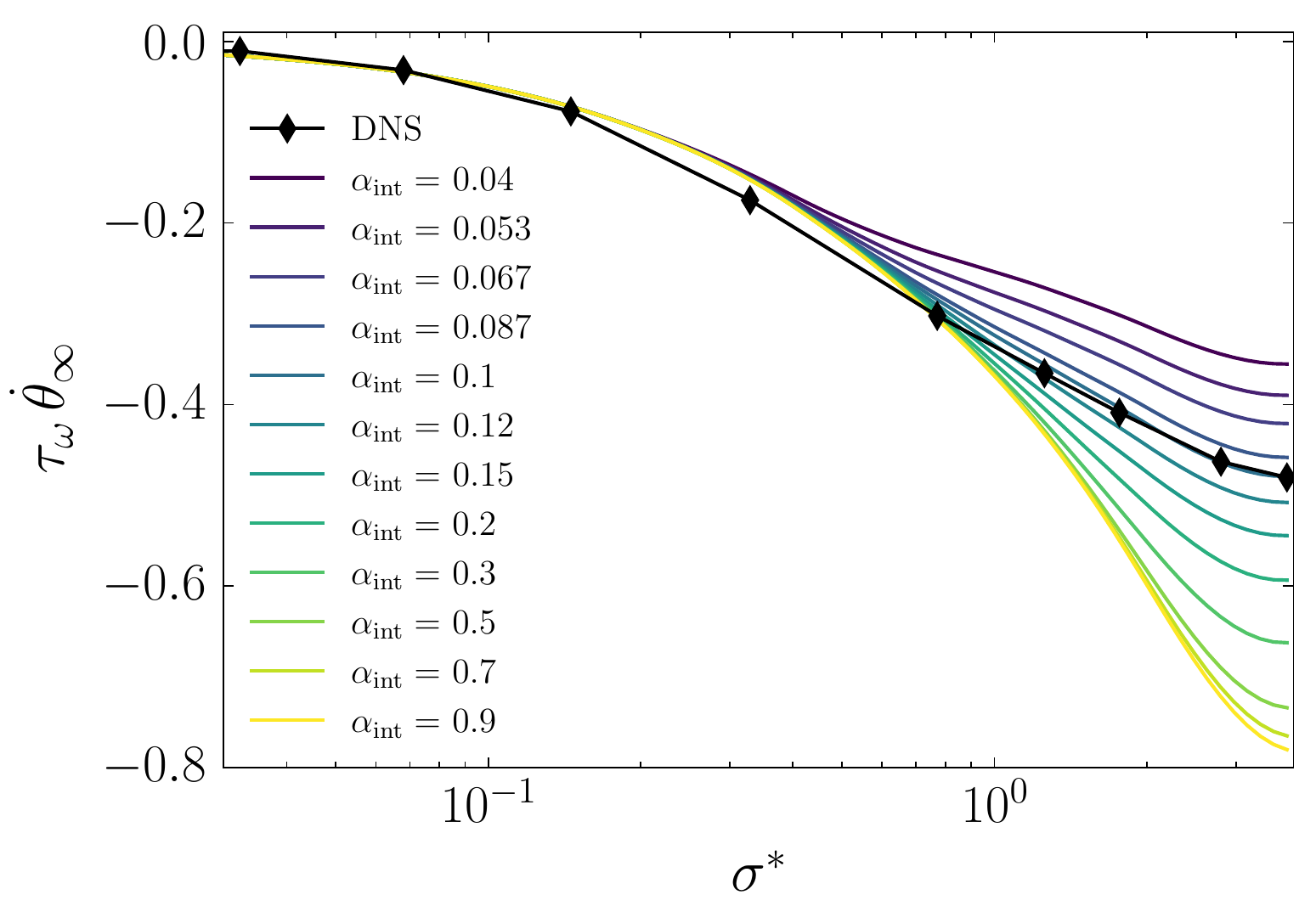}}
	\caption{\label{fig:tumbling_rate}
		(Colour online) \protect\subref{fig:Mean_DTheta} Average angular displacement of the unfolded angle $\delta\theta_t = \theta_t - \theta_0$ as a function of time for various values of shear rate  $\sigma^*$, showing a linear behaviour $\braket{\delta\theta_t} \simeq \dot{\theta}_\infty \, t$ at long times. DNS results are represented as solid lines and model values as dashed lines (here for $\alint=0.053$). 
		\protect\subref{fig:DThetaInf} Asymptotic angular velocity $\dot{\theta}_\infty$ as a function of 
		the shear rate parameter $\sigma^*$ from DNS ($\mathbin{\blacklozenge}$) and the model 
		(coloured lines) for various values of $\alint$, as labelled.}
\end{figure}

Using known properties of diffusions on the torus~\cite{bensoussan2011asymptotic}, one can show that the unfolded orientation angle $\theta_t\in\mathbb{R}$ associated to the stochastic model~\eqref{eq:folded_theta} follows exactly the same stochastic differential equation as its folded counterpart, but this time on $\mathbb{R}$ rather than on the periodic domain $[-\nicepi,\nicepi]$. One can thus write
\begin{equation}\label{eq:unfolded_theta}
	\delta \theta_t = \theta_t - \theta_0 = \int_0^t a(\theta_s)\, \dd s + \int_0^t b(\theta_s) \, \partial W_s,
\end{equation}
where $W_t$ is the exact same realisation of the Brownian motion as that entering the time evolution of $\btheta_t$. The drift and diffusion coefficients $a$ and $b$, defined in~\eqref{eq:coeff_sde} are periodic functions of $\theta_t$. It thus makes no difference evaluating them along the folded process $\btheta_t = \theta_t \; \mbox{mod}\; \pi$. For a sake of simplicity, let us assume that the initial orientation angle $\theta_0$ is chosen in $[-\nicepi,\nicepi]$, with a distribution given by the stationary solution $\Pst$ to the Fokker--Planck equation~\eqref{eq:FP_stationary}. This means that at any later time $t>0$, the folded angle $\btheta_t$ remains distributed according to the stationary law.  We then use It\^o's formulation of Eq.~\eqref{eq:unfolded_theta} to write its average as
\begin{align}
	\begin{aligned}
		\braket{\delta \theta_t}  =  \int_{0}^{t} \left\langle a(\btheta_s) 
		+ \frac{1}{4} \partial_{\btheta}b^2(\btheta_s) \right\rangle\dd s = t \left\langle a(\btheta)
		+\frac{1}{4} \partial_{\btheta}b^2(\btheta)\right\rangle.
	\end{aligned}
\end{align}
To simplify notations, expectations involving $\btheta$ are, here and in the sequel, understood as 
averages over the stationary distribution, and $\braket{a(\btheta)+\tfrac{1}{4} 
\partial_{\btheta} b^2 (\btheta)} \equiv \int_{-\nicepi}^{\nicepi} [a(\btheta)
+ \tfrac{1}{4} \partial_{\btheta}b^2(\btheta)] \, \pst(\btheta)\,\dd \btheta = -\pi\mathcal{J}$. We 
finally obtain that the average 
tumbling rate reads
\begin{align}\label{eq:tumbling_rate}
	\begin{aligned}
		\dot{\theta}_{\infty} \equiv 	\frac{\dd}{\dd t} \braket{\delta \theta_t}
		= -\pi\mathcal{J}.
	\end{aligned}
\end{align}
Figure~\ref{fig:tumbling_rate}\protect\subref{fig:Mean_DTheta} shows the time evolution of average angular displacements, both for the DNS (solid lines) and for the stochastic model with integral correlations (dashed lines), at various values of the mean shear rate. For the model, the tuning parameter was chosen to be $\alint=0.053$, corresponding to the optimal choice to reproduce both the average and the stationary distribution of the folded angle $\btheta$ (see Sec.~\ref{03subsec:correlations}). In both cases, the average angular displacement is zero when $\s^*=0$ and becomes increasingly negative when $\s^*$ increases.  For this specific choice of the parameter $\alint$, the model shows discrepancies with respect to DNS that are of the order of $20\%$ of the mean angular velocity $\dot{\theta}_{\infty}$ at the largest values of  $\sigma^*$.

In order to better assess the dependence upon the tuning parameter $\alint$ of the model, we show in  Fig.~\ref{fig:tumbling_rate}\protect\subref{fig:DThetaInf} how the average tumbling rate $\dot{\theta}_{\infty}$  changes when varying $\alint$ over an extended range.  DNS results are also shown as symbols and we find that the best match is obtained for $\alint\approx0.1$, a value that is roughly twice larger than that used to fit the stationary distribution in Sec.~\ref{03subsec:correlations}. This necessity to  increase significantly the tuning parameter, and consequently the strength of diffusion in the stochastic model, can be again interpreted as a signature of the coherent structures that are present in the actual turbulent flow.  For the stochastic model, the probability current, and consequently the average tumbling, only depends on how the degenerate stable fixed point $\btheta = 0$ is instantaneously perturbed. In the turbulent flow, $\btheta = 0$ is no more a fixed point for tracers that are captured in the persistent structures with a positive vorticity. They contribute to an increase of the average tumbling rate, but at the same time, they tend to uniformise the distribution of the folded angle. These two effects cannot be simultaneously captured by the stochastic model. They can only be reproduced individually using different values of the tuning parameter  $\alint$. Still, it is worth emphasising that, a single adjustment of  $\alint$ makes the model reproduce the tumbling rate $\dot{\theta}_\infty$ for all values of the shearing rates $\s^*$ that we considered.

\subsection{Variance of angular displacement}\label{03subsec:var_tumbling_rate}

The long-time diffusive behaviour of the orientation angle leads to introduce another tumbling rate, defined as the asymptotic growth rate of the variance of the angular displacement
\begin{equation} \label{eq:def_diff_coef}
	\DD=\lim_{t\to\infty} \frac{\dd}{\dd t} \left\langle\left(\delta \theta_t - \braket{\delta \theta_t}\right)^2\right\rangle.
\end{equation}
The analytical estimation of this diffusion coefficient, and thus of second-order statistics of the unfolded orientation angle $\theta_t$, is much less trivial than the calculation of the average angular displacement reported in previous subsection.
The first step consists in writing the second-order moment of the angular displacement as
$$
 \left\langle\left(\delta \theta_t - \braket{\delta \theta_t}\right)^2\right\rangle 
= \left\langle\left[{\textstyle \int_0^t   \left(\ha(\btheta_s)- \langle \ha(\btheta)\rangle\right)  \dd s}\right]^2\right\rangle  +  \left\langle{\textstyle \int_0^t b^2(\btheta_s) \,\dd s}\right\rangle,
$$
where $\ha(\theta) = a(\theta) +\tfrac{1}{4}\partial_{\theta}b^2(\theta)$   denotes the It\^o version of the model drift in \eqref{eq:unfolded_theta}. As in previous subsection, here again we assume that $\theta_0$ is chosen in $[-\nicepi,\nicepi]$ according to the stationary law $\pst$, and thus that the folded process $\btheta_t$ follows $\pst$ at any later time. This allows us writing
\begin{equation}\label{eq:contrib_var}
\frac{\dd}{\dd t} \left\langle\left(\delta \theta_t - \braket{\delta \theta_t}\right)^2\right\rangle =
2\left\langle \ha(\btheta_0)  \int_0^t  \left[\ha(\btheta_s)- \langle \ha(\btheta)\rangle\right] \dd s\right\rangle + \left\langle b^2(\btheta) \right\rangle. 
\end{equation}
To evaluate the first term on the right-hand side, we employ a useful method involving the auxiliary Poisson equation with an appropriate source term~\cite{mattingly2010}.  Let us consider $\eta$ solution for $\theta\in[-\nicepi,\nicepi]$ of
\begin{equation}\label{eq:poisson}
\frac{1}{2} b^2(\theta)\, \partial^2_{\theta}\eta(\theta) + \ha(\theta)\,\partial_{\theta}\eta(\theta) = \ha(\theta) - \braket{\ha(\btheta)}, \quad \text{with }
\partial_{\theta} \eta(-\nicepi) = \partial_{\theta}\eta(\nicepi).
\end{equation}
By Integrating \eqref{eq:poisson} against $\Pst$ and using the Fokker--Planck equation \eqref{eq:FP_stationary} for $\Pst$, one can show that the solution $\eta$ must be periodic as well.  
Applying the \Ito formula to $\eta(\btheta_t)$, evaluated along the folded process, we moreover get
$$
\begin{aligned}
\eta(\btheta_t) -\eta(\btheta_0) &=  \int_{0}^{t} \left[ \ha(\btheta_s)\,\partial_{\theta}\eta(\btheta_s) + \frac{1}{2} b^2(\btheta_s)\,\ha(\btheta_s)\,\partial^2_{\theta}\eta(\btheta_s) \right]\dd s + \int_0^t \partial_{\theta}\eta(\btheta_s)\,b(\btheta_s)\,\dd W_s \\
&= \int_{0}^{t} \left[\ha(\btheta_s)-\braket{\ha(\btheta)}\right] \dd s + \int_0^t \partial_{\theta}\eta(\btheta_s)\,b(\btheta_s)\,\dd W_s,
\end{aligned}
$$
where, for the second equality, we used  that $\eta$ solves the Poisson equation \eqref{eq:poisson}. Multiplying by $\ha(\btheta_0)$ and averaging with respect to noise, the contribution from the stochastic integral vanishes and
$$
\left\langle \ha(\btheta_0)  \int_0^t  \left[\ha(\btheta_s)-\braket{\ha(\btheta)}\right] \dd s \right\rangle  
=  \left\langle \ha(\btheta_0)\,\eta(\btheta_t) \right\rangle  -\left\langle \ha(\btheta)\,\eta(\btheta)\right\rangle.
$$
When $t\to\infty$, the first term de-correlates to approach 
$\braket{\ha(\btheta_0)}\braket{\eta(\btheta_t)}$.  Using both the Poisson equation for $\eta$ and 
the Fokker--Planck equation for $\Pst$, one can show that 
$ 2 \braket{\ha(\btheta)}\braket{\eta(\btheta)} - 2\braket{\ha(\btheta) \eta(\btheta)} =  
\braket{b^2(\btheta) (\partial_{\theta}\eta)^2(\btheta)}$.  Putting together the two contributions of 
\eqref{eq:contrib_var}, we finally find that the diffusion coefficient \eqref{eq:def_diff_coef} reads
\begin{align}\label{eq:var_tumb3}
\begin{aligned}
\DD =  \left\langle b^2(\btheta)\right\rangle +  \braket{b^2(\btheta) (\partial_{\theta}\eta)^2(\btheta)},
\end{aligned}
\end{align}
where we recall that averages involving $\btheta$ are taken with respect to the stationary 
distribution $\pst$.  This expression can be handle with precision using the semi-explicit solution of 
the Poisson equation: With the same notation  $\Psi(\theta)$ involved in the definition of the 
stationary distribution $\pst$ in \eqref{eq:PDF_stationary}, one has
\begin{align}
\partial_{\theta}\eta(\theta) &= 1 -  2\pi \mathcal{J} \frac{{\rm e}^{-\Psi(\theta)}}{b(\theta)} \left( \frac{
\int_{-\nicepi}^{\nicepi} 
\frac{\displaystyle  {\rm e}^{\Psi(\theta')} }{b(\theta')} \dd \theta'}{1 - {\rm e}^{\Psi(\nicepi)}} 
 -  \int_{-\nicepi}^\theta \frac{ {\rm e}^{\Psi(\theta')} }{b(\theta')} \dd\theta'\right). 
\end{align}
Note that when $\sigma^*\to0$, the source term $\braket{\ha(\btheta)}$ vanishes and the solution of~\eqref{eq:poisson} is trivially $\partial_{\theta}\eta(\theta)\equiv 1$. 

\begin{figure}[htbp]
\captionsetup[subfigure]{position=bottom, labelfont=bf,textfont=normalfont,	singlelinecheck=false,  justification=centering}
\centering
\subfloat[\label{fig:Var_DTheta_DNS}]{
\includegraphics[scale=0.48]{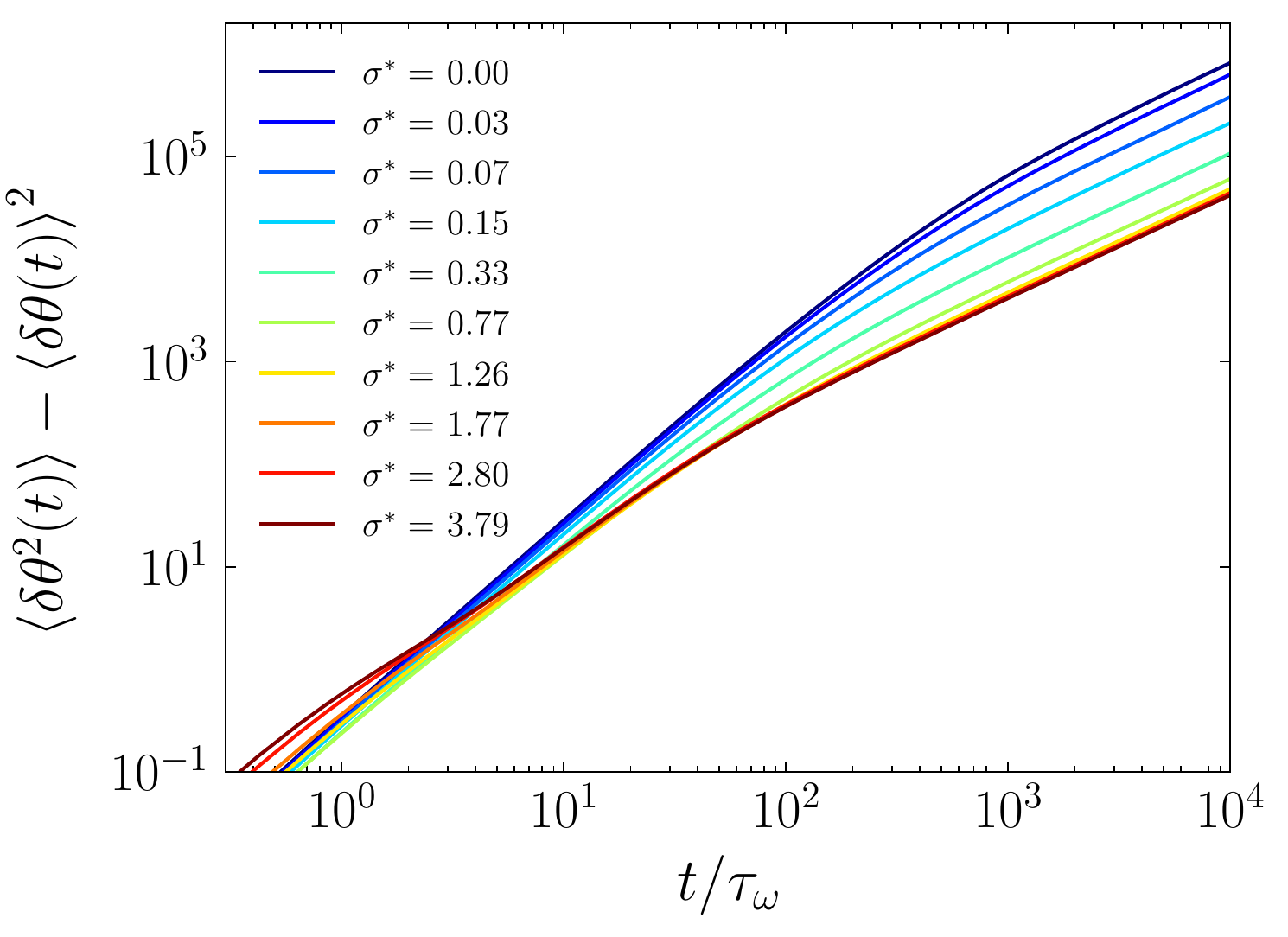}}
\hfil
\subfloat[\label{fig:D_Var_DTheta_Mdl_and_DNS}]{
\includegraphics[scale=0.48]{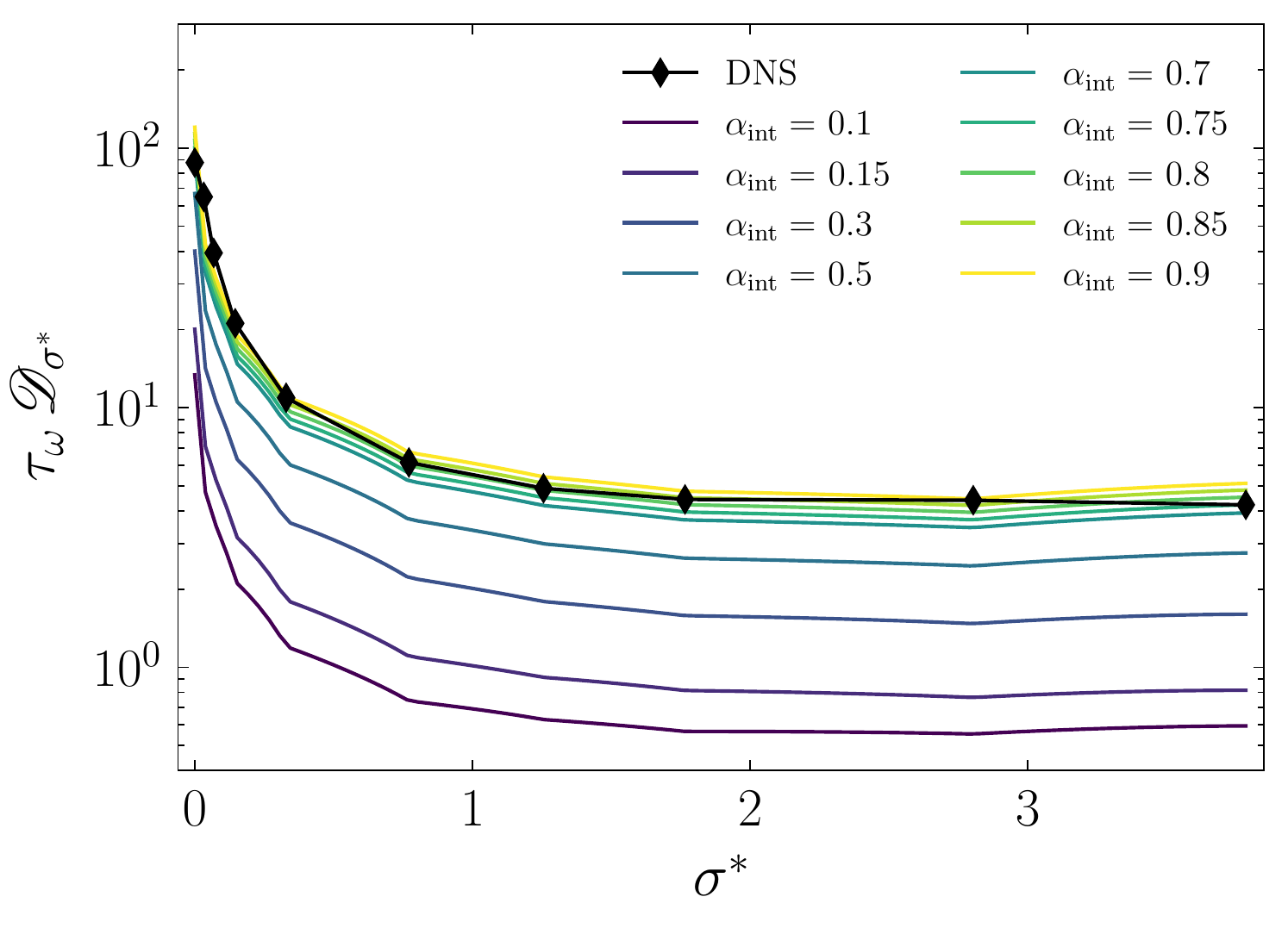}}
\caption{\label{fig:variance_dtheta} (Colour online)
\protect\subref{fig:Var_DTheta_DNS} Variance of the angular increment $\delta \theta_t$ as a function of time measured from DNS for various values of the shear rate. Data show a diffusive regime for $t\gtrsim 10^3\,\tau_\omega$. \protect\subref{fig:D_Var_DTheta_Mdl_and_DNS}  Diffusion coefficient $\DD$ as a function of $\sigma^*$ obtained from DNS for $2\times 10^3\le t/\tau_\omega \le 10^4$ ($\mathbin{\blacklozenge}$), and for the model (from Eq.~\protect\eqref{eq:var_tumb3}) for various values of $\alint$, as labelled.}
\end{figure}

Figure~\ref{fig:variance_dtheta}\protect\subref{fig:Var_DTheta_DNS}  shows DNS measurements of the variance of the angular 
increment $\delta\theta_t$ as a function of time. After a ballistic regime at very short times, there is 
a transition zone at intermediate times where the behaviour is possibly algebraic, with a  
super-diffusive exponent that seems to depend on the shear rate.  At much longer times, a diffusive 
behaviour is asymptotically reached. The time needed to reach the diffusive regime decreases with 
the shear rate. This behaviour can be explained in terms of the dependence upon $\s^*$ of the 
velocity gradient Lagrangian correlation time (see inset of Fig.~\ref{fig:sigma_correlations}\protect\subref{fig:CijklE_sigma}).
Figure~\ref{fig:variance_dtheta}\protect\subref{fig:Var_DTheta_DNS}  shows DNS measurements of the diffusion coefficient 
$\DD$ as function of the shear rate $\s^*$. When $\s^*$ increases, the diffusion coefficient 
becomes smaller because shear tends to deplete the structures of the flow. For the three largest 
values of shear, $\DD$ becomes almost constant. This could originate from the fact that our 
simulations are performed in a finite-size domain. Indeed, when $\s^*$ is large,  the elongated 
structures of the flow start to get influenced by periodic boundary conditions and align with the 
direction of shear. This geometrical effect could affect our measurements. 

The diffusion coefficient $\DD$ obtained from Eq.~\eqref{eq:var_tumb3} for the model is also shown 
in Fig.~\ref{fig:variance_dtheta}\protect\subref{fig:D_Var_DTheta_Mdl_and_DNS} for different values of $\alint$. Its overall dependence upon 
$\s^*$ is very similar to DNS data. However, one observes that the value $\alint = 0.1$ of the fitting 
parameter that was previously used to reproduce the average angular displacement 
(Fig.~\ref{fig:tumbling_rate}\protect\subref{fig:DThetaInf} largely underestimates its variance. Quantitative agreements 
indeed require to choose $\alint\approx 0.8$.  The two statistics can thus not be reproduced with the same value of the fitting parameter. 
The needed increase of $\alint$ to fit second-order statistics 
can be interpreted as a way for the stochastic model to compensate fat tails that develop in the 
distribution of angular increments, so that the contribution of large fluctuations is accounted for by a larger diffusion.

\begin{figure}[t!]
\captionsetup[subfigure]{position=bottom, labelfont=bf,textfont=normalfont,	singlelinecheck=false,  justification=centering}
\centering
\subfloat[\label{fig:PDF_DTheta_s0}]{
\includegraphics[scale=0.46]{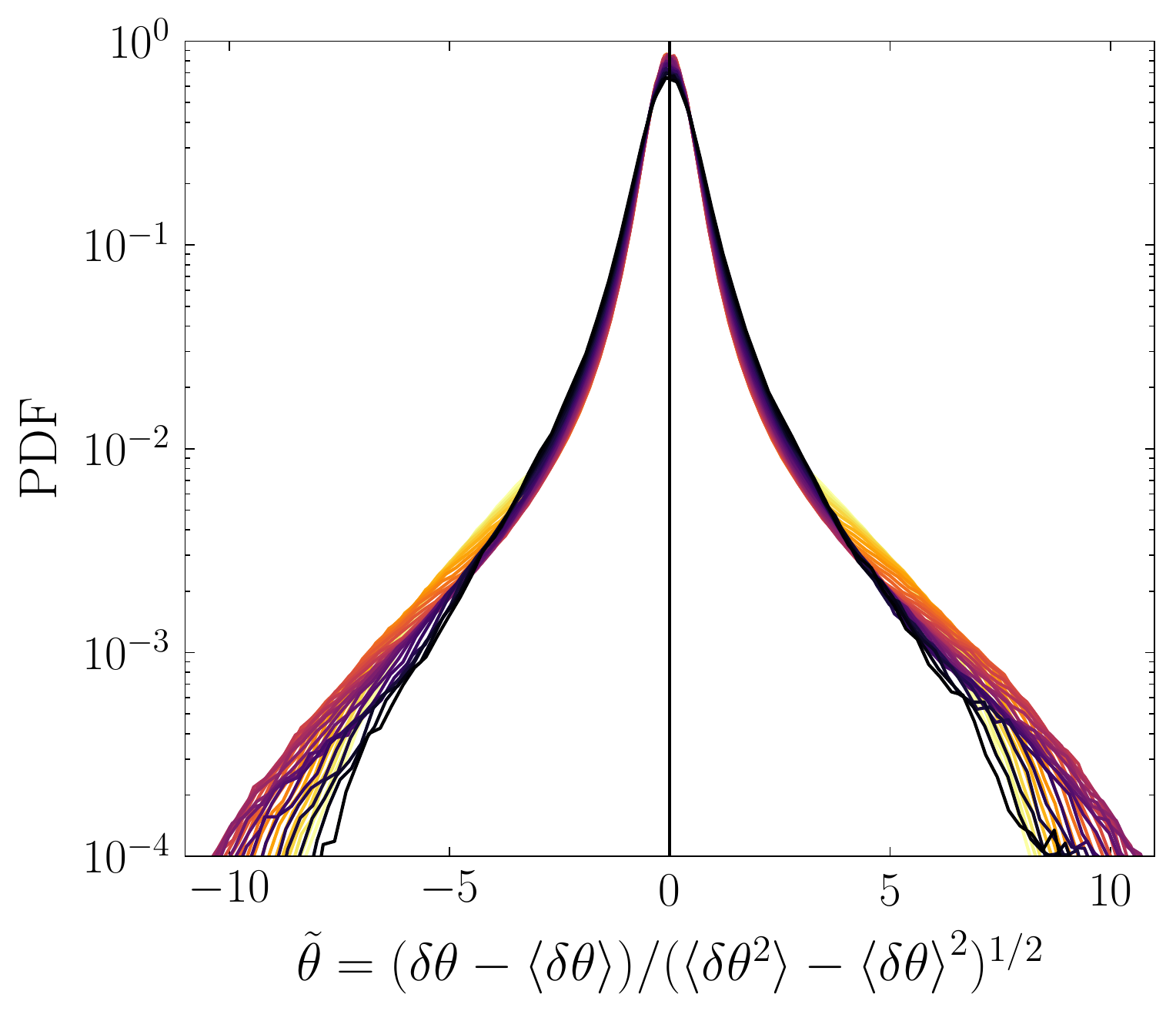}}
\hfil
\subfloat[\label{fig:PDF_DTheta_s8}]{
\includegraphics[scale=0.46]{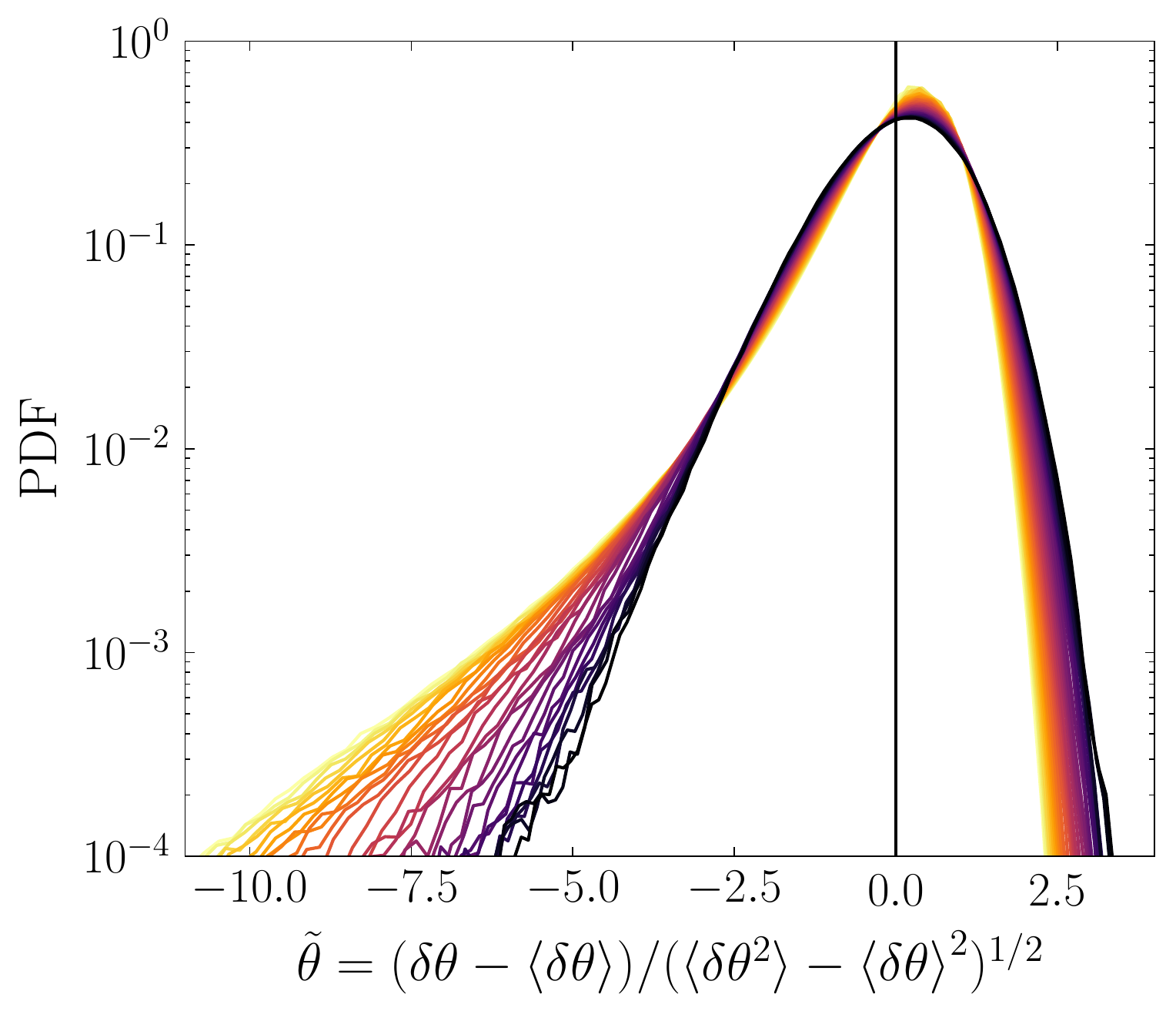}}
\caption{\label{fig:pdf_dtheta} (Colour online)
Probability density function (PDF) of the angular increment $\delta \theta_t = \theta_t-\theta_0$ 
	centred and rescaled to unit variance for $\sigma^*=0$ \protect\subref{fig:PDF_DTheta_s0}, and $\sigma^*= 2.8$ \protect\subref{fig:PDF_DTheta_s8}. 
	The different curves correspond to various time lags logarithmically spaced between 
	$t=80\,\tau_\omega$ and $t=6000\,\tau_\omega$ (from light yellow to dark violet).}
\end{figure}

The violent fluctuations of angular increments are evidenced in Fig.~\ref{fig:pdf_dtheta}, which 
shows the probability density function of $\delta\theta_t$ obtained from DNS at different times lags 
$t$ and for $\s^*=0$ \protect\subref{fig:PDF_DTheta_s0} and $\s^*=2.8$ \protect\subref{fig:PDF_DTheta_s8}.  At first glance, one observes a 
strong qualitative difference between these two cases: Shear completely depletes fluctuations at 
positive values, as already observed for vorticity (Fig.~\ref{fig:stat_fluids}\protect\subref{fig:stat_fluids_b}).  
Still the two cases share common features. First, as the time lag increases, the tails of both distributions tend to shrink 
and possibly approach a Gaussian behaviour. This is more evident for $\s^*>0$ that shows an 
important depletion of large fluctuations when $t$ increases.
Second, one observes in both cases intermediate asymptotic tails (for angular increments from 
2 to 10 times their standard deviation) where the probability density decreases slower than an 
exponential. 
We interpret this behaviour as a contribution from the intermittent, quasi ballistic excursions of the 
rods orientation observed in Fig.~\ref{fig:traj_Dtheta} and that are due to trapping by coherent 
vortical structures.
These violent events should have a visible signature on the moments of the angular increment, 
possibly explaining why the model cannot reproduce both the mean and the variance with the same 
amplitude factor $\alint$. Accounting for such effects would require to change paradigm and to 
consider jumps and not pure diffusion in the rotation dynamics of rods.

\section{Conclusions}\label{sec:conclusion}
To summarise, we have here studied the dynamics and statistics of rod-like particles transported by a homogeneous turbulent two-dimensional shear flow.  We have introduced a Lagrangian stochastic model that is able to reproduce the combined effects of an average shear and of turbulent fluctuations on the orientation dynamics of such particles. The modelled velocity gradient is the superposition of a deterministic, stationary part and of random, delta-correlated-in-time component. This allowed us to derive analytical predictions for the statistics of the rods orientation, which were assessed by comparison to the results of direct numerical simulations. 

We have first focused on single-time stationary statistics of the rods orientation. The model allowed us writing an explicit expression for the stationary probability density of the folded angle $\bar{\theta}$ in term of the solutions with a constant probability current of the associated Fokker--Planck equation.  The dependence of this distribution upon the shear rate parameter $\s^*$  is also explicit and eases comparison with DNS results. Three different expressions for the model fluid-gradient correlations were tested and our results reveals that orientation statistics are well reproduced only if both anisotropies and effects of long-time correlations of the Lagrangian velocity gradients are taken into account. Fine-tuning the model to fit DNS moreover requires to properly choose the value of a dimensionless parameter $\alint$, which controls the overall amplitude of velocity gradient fluctuations.

We then turned to two-time statistics and studied the evolution of the rods' unfolded orientation 
over a given time lag. This led us to introduce two new definitions of the ``tumbling rate''. The first, 
which solely applies in the presence of shear, is given by the asymptotic average angular velocity of 
the rods. It is directly proportional to the constant probability current of the stationary distribution 
and actually measures by how many turns per unit time the rods rotate under the influence of the 
mean shear. The second definition of a tumbling rate has the advantage to also apply in the absence 
of average shear. It is equal to the diffusion coefficient of the rods' angular increment. These two 
tumbling rates can be derived analytically for the model. However, when comparing their values with 
DNS, one finds that they both require increasingly large values of the fitting fluctuation amplitude 
$\alint$. In short, our stochastic model is found to entail most important qualitative aspects of the 
orientation dynamics but reproducing quantitatively several high-order statistics cannot be done 
with a single value the fitting parameter.

We interpret such pitfalls as originating from rare, violent events in the orientation dynamics that 
occur when the rods are trapped in coherent vortical structures for extremely long times. Our model 
only accounts for the time-averaged contribution of such events, since it makes use of Lagrangian 
integral correlation times of fluid-velocity gradients. DNS results suggest that this might not be 
enough to reproduce the full distribution of angular increments. Nevertheless, to our view, this issue 
actually opens new modelling perspectives. The trajectories of the angular increment $\delta 
\theta_t$ are a combination of diffusion periods and long-range excursions, that strongly bring 
to mind known behaviours of L\'{e}vy processes. In this respect, further developments could consist 
in modelling the fluctuations of the fluid-velocity gradient in terms of a jump process, in an 
attempt to get a more realistic description of DNS. Such processes are relatively straightforward to 
simulate. However, their use requires to introduce additional parameters that need to be calibrated, 
adding another level of complexity.

Finally let us stress that many aspects discussed in this article, and in particular the stochastic model 
that we introduce, straightforwardly extend to three-dimensional situations. The physics is however 
expected to be much richer, because different behaviours will arise depending whether the particles 
are rod or disk-shaped. We still expect to observe again non-trivial contributions from flow 
structures consisting of vortex tubes that are preferentially aligned with the span-wise 
direction~\cite{pumir1996turbulence}.
Modelling the orientation dynamics of anisotropic particles in three dimensions becomes however 
much harder, in particular to account for the intricate alignment of the Cauchy--Green tensor with 
both passive vectors and vorticity~\cite{ni2014alignment}.  Also, tracking the cumulative rotation of 
rods becomes more complicated than in two dimensions;  Nonetheless, first steps towards a 
three-dimensional extension of the unfolded angular dynamics have been made in~\cite{campana}. 
All of these developments are subject to ongoing work.

\begin{acknowledgments}
The authors are grateful to the OPAL infrastructure from Universit\'{e} C\^{o}te d'Azur for providing 
computational resources and support. We benefited from stimulating discussions with Christophe 
Henry and Dario Vincenzi who are warmly acknowledged. This work has been supported by EDF 
R\&D (projects PTHL of MFEE and VERONA of LNHE) and by the French government, through the 
Investments for the Future project UCAJEDI ANR-15-IDEX-01 (grant no.~ANR-21-CE30-0040-01) 
managed by the Agence Nationale de la Recherche.
\end{acknowledgments}


\appendix

\section{\Ito's lemma on the orientation}\label{apndx:2d_ito_lemma}
We recall first the relation $\B_{ijnm}\B_{klnm}=2\D_{ijkl}$, and introduce the notation: 
\begin{equation}\label{eq:general_tensor_incomp}
	\D_{ijkl} =
	\begingroup
	\renewcommand*{\arraystretch}{1.2}
	\begin{pmatrix}
		\D_{1111} & \D_{1112} & \D_{1121} & \D_{1122} \\
		\D_{1211} & \D_{1212} & \D_{1221} & \D_{1222} \\
		\D_{2111} & \D_{2112} & \D_{2121} & \D_{2122} \\
		\D_{2211} & \D_{2212} & \D_{2221} & \D_{2222} \\
	\end{pmatrix}
	\endgroup 
	=
	\begin{pmatrix}
		f  			&  h  &  j 		& -f \\
					&  g  &  k		& -h \\
		\text{sym.} &  	  &  \ell	& -j \\
					& 	  & 		&  f
	\end{pmatrix}.
\end{equation}
We derive the dynamics of the orientation angle $\btheta=\arctan(\nicefrac{r_2}{r_1})$, applying the \Ito's Lemma to  the separation vector in \eqref{eq:separation_model} ($d=2$), that we rewrite first in its \Ito form
\begin{align*}
dr_i  &= \braket{A_{ij}} r_j dt +  \D_{jlij} r_l dt 
+   \B_{ilk} r_l  d\w^k_t,
\end{align*}
using the renumbering $\sum_{n,m=1}^2 \B_{ilnm} \partial \mathbb{W}_{nm} = \sum_{k=1}^4 \B_{ilk} \partial \w^{k}$, 
The multidimensional Ito's lemma applied to $\arctan(\nicefrac{r_2}{r_1})$ gives
\begin{equation*}
\begin{aligned}
		d\btheta =& \sum_{i,j}^{2} J^{\btheta}_i \braket{A_{ij}} r_j dt
		+ \sum_{i,j,l} J^{\btheta}_i \D_{jlij} r_l  dt +\frac{1}{2} \sum_{i,j,l,l'}^{2} \sum_{k}^{4} H^{\btheta}_{ij} 	\left(\B_{ilk} r_l \B_{jl'k} r_{l'}\right) dt
		+\sum_{i,l}^{2} \sum_{k}^{4} J^{\btheta}_i \B_{ilk} r_l  d\w^k_t;
\end{aligned}
\end{equation*}
where $J^{\btheta}_i=-\sum_{j}^{2}\varepsilon_{ij} r_j 
\|r\|^{-2}$ and $H^{\btheta}_{ij}
=\sum_{l,l'}^{2} \left(\varepsilon_{il} r_l r_j 
+ \varepsilon_{jl'} r_{l'} r_i \right) \|r\|^{-4}$ are respectively the gradient and 
Hessian matrix of $\arctan(\nicefrac{r_2}{r_1})$, denoting with $\varepsilon_{ij}$ the 2D-Levi-Civita 
symbol, 
$\varepsilon_{ij}={\tiny{\begin{pmatrix}0&1\\-1&0\end{pmatrix}}}.$
In the equation above, the stochastic integral term can be reduced (with the help of the Brownian martingale representation -- see e.g \citet[Theorem 4.2,Chpt 3]{karatzas2012brownian})  to a stochastic  term (equivalent in law only)  driven by a single Brownian motion $(W_t)$, weighted  with the diffusion  matrix norm: 
\begin{equation}\label{eq:sde_theta}
	\begin{aligned}
		d\btheta
		\overset{\textrm{law}}{=}& 
		\Big( \sum_{i,j}^{2} J^{\btheta}_i \braket{A_{ij}} r_j dt
		+ \sum_{i,j,l} J^{\btheta}_i \D_{jlij} r_l  dt 
		+\frac{1}{2} \sum_{i,j,l,l'}^{2} \sum_{k}^{4} H^{\btheta}_{ij} 	\left(\B_{ilk} r_l \B_{jl'k} r_{l'}\right) \Big)  dt \\
		&  +\Big( \sum_{k}^{4}  \big( \sum_{i,l}^2 J^{\btheta}_i  \B_{ilk} r_l \big)^2\; \Big)^{\frac{1}{2}} 
		dW_t\\
		{=}& \ (\RN{1}+ \RN{2} + \RN{3}) \ dt  + \RN{4} \ dW_t.
	\end{aligned}
\end{equation}
We detail separately the computation of the four terms above.

\paragraph*{Drift term $\RN{1}$. } Introducing $p_i=r_i /\|r\|$, i.e.  $\bm p = (\cos\btheta,\sin\btheta)$, and since the only component different from zero of $\braket{A_{ij}}$ is $\braket{A_{12}}=\sigma^*$, 
\begin{align*}
	\RN{1} = \sum_{i,j}^{2} J^{\btheta}_i \braket{A_{ij}} r_j 
	&= -\sum_{i\neq m, j}^{2} \varepsilon_{im} p_m \braket{A_{ij}} p_j = -\sigma^* p_2^2 dt = \frac{\sigma^*}{2}\left(\cos(2\btheta)-1\right). 
\end{align*}

\paragraph*{Drift term $\RN{2}$. } This contribution is identified to be zero: using again the definition of $ J^{\btheta}_i$, and introducing the symmetries in  matrix~\eqref{eq:general_tensor_incomp}
\begin{align*}
\RN{2} & =  - \sum_{i\neq \ell}^{2}\varepsilon_{i\ell} p_\ell  \sum_{j,l}^{2} \D_{jlij} p_l =  - 
\sum_{j,l,}^{2} (\D_{jl1j}p_2 p_l - \D_{jl2j}p_1 p_l ) = 0.
\end{align*}

\paragraph*{Drift term $\RN{3}$. } Developing the computation with the symmetries in  matrix~\eqref{eq:general_tensor_incomp}:
\begin{align*}
\RN{3} & =
\sum_{i,j,l,l'}^{2}  H^{\btheta}_{ij} \D_{iljl'} \ r_l r_{l'} = \sum_{i \neq m}^{2} \sum_{j \neq n}^{2} \sum_{l,l'}^{2}  \D_{iljl'} \left(\varepsilon_{im} p_m 
	p_j+ \varepsilon_{jn} p_n p_i\right) p_l p_{l'} \\
	&=2\sum_{l,l'}^{2}\left(
	\left(\D_{1l1l'}-\D_{2l2l'}\right)p_1p_2
	+\frac{1}{2}\left(\D_{1l2l'}+\D_{2l1l'}\right)
	\left(p_2^2-p_1^2\right)\right) p_l p_{l'}\\
	& =2  \Big(3(j+h) p_1^2 p_2^2 -j p_1^4 -h p_2^4
	+(2f -\ell -k) p_1^3 p_2 +(-2f +g +k) p_2^3 p_1 \Big).
\end{align*}
Next, introducing the following equalities from Cartesian to polar coordinates
\begin{align*}
	& p_1^3p_2 =\tfrac{1}{4} (\sin(2\btheta) +\tfrac{1}{2}\sin(4\btheta)),  
	& p_1^2p_2^2 = \tfrac{1}{8} (1 -\cos(4\btheta)), \qquad ~~~
	& p_1^4 =\tfrac{1}{8} (3 +4\cos(2\btheta) +\cos(4\btheta)), \\ 		
	&p_2^3p_1 =\tfrac{1}{4} (\sin(2\btheta) -\tfrac{1}{2}\sin(4\btheta)), 
	& 
	& p_2^4 = \tfrac{1}{8} (3 -4\cos(2\btheta) +\cos(4\btheta)),
\end{align*}
\begin{align*}
	\RN{3} =& 
	-\frac{\gamma_3}{2} \sin(2\btheta) -\gamma_4 \sin(4\btheta) 
	+\frac{\gamma_1}{2} \cos(2\btheta) +\gamma_2 \cos(4\btheta),
\end{align*}
with  
\begin{equation}\label{eq:apendix-defgammas}
\begin{aligned}
&\gamma_1 = 2(h - j) & &\gamma_2 = -h - j  \\
&\gamma_3 = \ell - g & &\gamma_4 = -f + \tfrac{1}{2}k + \tfrac{1}{4}(\ell + g).
\end{aligned}	
\end{equation}

\paragraph*{Diffusion term $\RN{4}$. } We denoted this coefficient $b(\bm{p})$, with $b^2(\bm{p}) =  \sum_{k}^{4}  \Big( \sum_{i,l} J^{\btheta}_i  \B_{ilk} r_l \Big)^2$. Using  $(J^{\btheta}_1 r_l,J^{\btheta}_2 r_l) = (- p_2p_l, p_1 p_l)$
\begin{align*}
b^2(p) =  \sum_{k}^{4} \left( - p_2p_1  \B_{11k} + p_1^2 \B_{21k}   -  p_2^2  \B_{12k}  + p_1 p_2 \B_{22k} \right)^2.
\end{align*} 
Denoting $\B_{ijk}$ as the $4\times 4$ matrix  $\mathbb{B}=B_{ 2(i-1)+j,k}$, and $v(p)$ the $\mathbb{R}^4$ element $(-p_1 p_2, - p_2^2, p_1^2 , p_1 p_2)$, we obtain 
\begin{align}\label{eq:app_b2}
\begin{aligned}
b^2(p) &  =  \| \mathbb{B} v(p) \|^2  = v(p)^t \mathbb{B}^t \mathbb{B} v(p)  = 2 v(p)^t \mathbb{D} v(p) 
 = 2 \{(4f - 2k) p_1^2 p_2^2  + 4 h p_2^3 p_1  - 4 j p_1^3 p_2   +  \ell p_1^4 + g p_2^4  \}.
\end{aligned}
\end{align} 
Using next the polar coordinates relations introduced  earlier 
\begin{align*}
	b^2(\btheta)
	&= \gamma_0 +\gamma_1 \sin(2\btheta) +\gamma_2 \sin(4\btheta) 
	+\gamma_3 \cos(2\btheta) +\gamma_4 \cos(4\btheta),
\end{align*}
where $\gamma_0 = f +\tfrac{1}{2}k +\tfrac{3}{4}(\ell+g)$.  We report in Table \ref{tab:gamma_values}, the measured values of $\gamma_i$, for $\D^{\text{aniso}}(\sigma^*)$ and $\D^{\text{int}}(\sigma^*)$. 
{ 
	\renewcommand{\arraystretch}{1.4}
	\begin{table}[h!]
		\centering
	
\begin{tabular}{llrrrrrrrrrr}
	\hline
	$\sigma^*$ &  
	& $0.00$ & $0.03$ & $0.07$ & $0.15$ & $0.33$ & $0.77$ & $1.26$ & $1.77$ & $2.80$ & 
	$3.79$ \\
	\hline
	\multirow{2}{*}{$\gamma_0$} &
	\multicolumn{1}{c}{$\Daniso$} & 
	\multicolumn{1}{r}{$0.7565$} &  \multicolumn{1}{r}{$0.7559$} & 
	\multicolumn{1}{r}{$0.7612$} &  \multicolumn{1}{r}{$0.7433$} &  
	\multicolumn{1}{r}{$0.7311$} &  \multicolumn{1}{r}{$0.7084$} &  
	\multicolumn{1}{r}{$0.7210$} &  \multicolumn{1}{r}{$0.7223$} &  
	\multicolumn{1}{r}{$0.7065$} &  \multicolumn{1}{r}{$0.7389$} \\
	\multicolumn{1}{l}{} &
	\multicolumn{1}{c}{$\Dint$} & 
	\multicolumn{1}{r}{$66.8319$} & \multicolumn{1}{r}{$49.3522$} & 
	\multicolumn{1}{r}{$36.7193$} & \multicolumn{1}{r}{$21.4299$} & 
	\multicolumn{1}{r}{$12.3579$} & \multicolumn{1}{r}{$ 7.7984$} &  
	\multicolumn{1}{r}{$ 6.5343$} & \multicolumn{1}{r}{$ 6.0211$} &  
	\multicolumn{1}{r}{$ 6.0748$} & \multicolumn{1}{r}{$ 8.0637$} \\
	\hline
	\multirow{2}{*}{$\gamma_1$} &
	\multicolumn{1}{c}{$\Daniso$} & 
	\multicolumn{1}{r}{$ 0.0033$} & \multicolumn{1}{r}{$-0.0008$} & 
	\multicolumn{1}{r}{$-0.0160$} & \multicolumn{1}{r}{$-0.0734$} & 
	\multicolumn{1}{r}{$-0.1642$} & \multicolumn{1}{r}{$-0.1883$} & 
	\multicolumn{1}{r}{$-0.1665$} & \multicolumn{1}{r}{$-0.1518$} & 
	\multicolumn{1}{r}{$-0.1165$} & \multicolumn{1}{r}{$-0.1007$} \\
	\multicolumn{1}{c}{} &
	\multicolumn{1}{c}{$\Dint$} & 
	\multicolumn{1}{r}{$ 0.6818$} & \multicolumn{1}{r}{$0.2007$} &  
	\multicolumn{1}{r}{$ 0.2778$} & \multicolumn{1}{r}{$-0.6735$} & 
	\multicolumn{1}{r}{$-2.4257$} & \multicolumn{1}{r}{$-1.3606$} & 
	\multicolumn{1}{r}{$-0.7794$} & \multicolumn{1}{r}{$-0.6748$} & 
	\multicolumn{1}{r}{$-0.4338$} & \multicolumn{1}{r}{$-0.3979$} \\
	\hline
	\multirow{2}{*}{$\gamma_2$} &
	\multicolumn{1}{c}{$\Daniso$} & 
	\multicolumn{1}{r}{$-0.0015$} &  \multicolumn{1}{r}{$-0.0002$} & 
	\multicolumn{1}{r}{$-0.0022$} &  \multicolumn{1}{r}{$-0.0027$} &  
	\multicolumn{1}{r}{$ 0.0052$} &  \multicolumn{1}{r}{$ 0.0317$} &  
	\multicolumn{1}{r}{$ 0.0442$} &  \multicolumn{1}{r}{$ 0.0455$} &  
	\multicolumn{1}{r}{$ 0.0394$} &  \multicolumn{1}{r}{$ 0.0354$} \\
	\multicolumn{1}{l}{} &
	\multicolumn{1}{c}{$\Dint$} & 
	\multicolumn{1}{r}{$ 0.0034$} &  \multicolumn{1}{r}{$ 0.0184$} & 
	\multicolumn{1}{r}{$-0.0002$} &  \multicolumn{1}{r}{$-0.0171$} &  
	\multicolumn{1}{r}{$ 0.0581$} &  \multicolumn{1}{r}{$ 0.2195$} &  
	\multicolumn{1}{r}{$ 0.1815$} &  \multicolumn{1}{r}{$ 0.1779$} &  
	\multicolumn{1}{r}{$ 0.1428$} &  \multicolumn{1}{r}{$ 0.1494$} \\
	\hline
	\multirow{2}{*}{$\gamma_3$} &
	\multicolumn{1}{c}{$\Daniso$} & 
	\multicolumn{1}{r}{$-0.0095$} & \multicolumn{1}{r}{$-0.0014$} & 
	\multicolumn{1}{r}{$-0.0032$} & \multicolumn{1}{r}{$-0.0179$} & 
	\multicolumn{1}{r}{$-0.1031$} & \multicolumn{1}{r}{$-0.2961$} & 
	\multicolumn{1}{r}{$-0.4134$} & \multicolumn{1}{r}{$-0.4818$} & 
	\multicolumn{1}{r}{$-0.5741$} & \multicolumn{1}{r}{$-0.7004$} \\
	\multicolumn{1}{l}{} &
	\multicolumn{1}{c}{$\Dint$} & 
	\multicolumn{1}{r}{$-0.1937$} & \multicolumn{1}{r}{$ 0.0441$} & 
	\multicolumn{1}{r}{$-0.2509$} & \multicolumn{1}{r}{$-0.5446$} & 
	\multicolumn{1}{r}{$-2.1580$} & \multicolumn{1}{r}{$-3.6875$} & 
	\multicolumn{1}{r}{$-4.4623$} & \multicolumn{1}{r}{$-4.9191$} & 
	\multicolumn{1}{r}{$-5.9903$} & \multicolumn{1}{r}{$-9.2379$} \\
	\hline
	\multirow{2}{*}{$\gamma_4$} &
	\multicolumn{1}{c}{$\Daniso$} & 
	\multicolumn{1}{r}{$ 0.0028$} & \multicolumn{1}{r}{$ 0.0021$} &  
	\multicolumn{1}{r}{$ 0.0004$} & \multicolumn{1}{r}{$-0.0004$} & 
	\multicolumn{1}{r}{$-0.0057$} & \multicolumn{1}{r}{$ 0.0043$} &  
	\multicolumn{1}{r}{$ 0.0328$} & \multicolumn{1}{r}{$ 0.0528$} &  
	\multicolumn{1}{l}{$ 0.0885$} & \multicolumn{1}{r}{$ 0.1300$} \\	
	\multicolumn{1}{l}{} &
	\multicolumn{1}{c}{$\Dint$} & 
	\multicolumn{1}{r}{$ 0.0477$} & \multicolumn{1}{r}{$0.0058$} &  
	\multicolumn{1}{r}{$ 0.0460$} & \multicolumn{1}{r}{$0.0235$} & 
	\multicolumn{1}{r}{$-0.0264$} & \multicolumn{1}{r}{$0.2027$} &  
	\multicolumn{1}{r}{$ 0.4348$} & \multicolumn{1}{r}{$0.5986$} &  
	\multicolumn{1}{r}{$ 0.9369$} & \multicolumn{1}{r}{$1.7944$} \\
	\hline
\end{tabular}

	\caption{\label{tab:gamma_values}
		Values of $\gamma_i$ (with $i=0,\ldots,4$) used in the model (in the normalised case, \ie by 
		multiplying for $\tauO$) when we use $\Daniso $ or $\Dint$ to study the model as a function of 
		the shear rate $\s^*$. Here, the values of the tuning parameters are $\alaniso=\alint=1$.}
	\end{table}
}

\section{Strict positivity of the folded angle diffusion coefficient $b$}\label{apndx:strict_pos_b}
The strict positivity of $b$ is required to develop analytical expressions through the stochastic approach. We discuss here the three cases studied in this paper. To facilitate the discussion, we start from the expression \eqref{eq:app_b2} for $b^2$ expressed with the Cartesian coordinates. 
\paragraph{The $\D^{\text{iso}}$ case} simplifies \eqref{eq:app_b2} with  $f=-k=$, $g=l=3f$, $h=j=0$, and so   
\begin{align*}
b^2(p) &  = 6f p_1^2 p_2^2  + 3 f  p_1^4 +  3 f  p_2^4 = 3 f > 0.
\end{align*} 

\paragraph{The $\D^{\text{aniso}}$ case} (see Fig. \ref{fig:sigma_correlations}\protect\subref{fig:Cijkl0_sigma}) allows  to minorate \eqref{eq:app_b2} with $0\leq f = -k $, $0 \leq f \leq l \leq g $,  $0\leq j \leq f$ and $ k\leq h \leq 0 $.  Assuming first $|p_1| \leq  |p_2|$, then  $4 h p_2^3 p_1 > 4 h p_2^4$, $- 4 j p_1^3 p_2 > - 4 j p_1^2 p_2^2$, and 
\begin{align*}
b^2(p)  =  6f p_1^2 p_2^2  + 4 h p_2^3 p_1  - 4 j p_1^3 p_2   + g p_2^4 +  l p_1^4   & \geq 6f p_1^2 p_2^2  +  4 h p_2^4  - 4 j p_1^2 p_2^2 +   g p_2^4 +  l p_1^4  \\
& = (6f - 4j)  p_1^2 p_2^2  + (g +4 h) p_2^4  +  l p_1^4 \geq \min \{ (6f - 4j)/2 ,(g +4 h), l \}. 
\end{align*} 
Assuming  next $|p_1| \geq  |p_2|$,  we  also consider that $(l + 4 h) > 0 $ in Fig. \ref{fig:sigma_correlations}\protect\subref{fig:Cijkl0_sigma}. Then 
\begin{align*}
b^2(p)  = 6f  p_1^2 p_2^2  + 4 h p_2^3 p_1  - 4 j p_1^3 p_2   + g p_2^4 +  l p_1^4 & \geq 6f p_1^2 p_2^2  +  4 h p_2^2 p_1^2  - 4 j p_1^4  +   g p_2^4 +  l p_1^4  \\
&= (6f + 4h)  p_1^2 p_2^2  + (l - 4j) p_1^4  +  g p_2^4 \geq \min \{ (6f + 4h)/2 ,(l - 4 j), g\}. 
\end{align*} 
So for any $p$,
\begin{align*}
b^2(p) & \geq \min \{ (6f - 4j)/2 ,(g +4 h), l \} \wedge \min \{ (6f + 4h)/2 ,(l -4 j), g\}.
\end{align*}
According to Figure \ref{fig:sigma_correlations}\protect\subref{fig:Cijkl0_sigma}, the smallest term above is $(l-4j)>0$. The diffusion is then strictly positive for the considered values of shear.

\paragraph{The $\D^{\text{int}}$ case} (see Fig. \ref{fig:sigma_correlations}\protect\subref{fig:CijklE_sigma})  simplifies \eqref{eq:app_b2} with  $f = h =j =0$, $k\leq 0$, $g>l>0$ to
\begin{align*}
b^2(p) = - 2k p_1^2 p_2^2 + g p_2^4 +  l p_1^4  = l  - 2(l+k) p_1^2 p_2^2  + (g - l) p_2^4 > 0. 
\end{align*} 
The diffusion is then again strictly positive for the considered values of shear.

%

\end{document}